\documentstyle[12pt]{article}

\newcommand{\bge}{\begin{equation}}
\newcommand{\ee}{\end{equation}}
\newcommand{\mixten}[3]{{#1}^{#2}_{\phantom{{#2}} #3}}

\newcommand{\startappendix}{
\renewcommand{\thesection}{\Alph{section}}}

\begin{document}
\renewcommand{\thepage}{ }
 \begin{flushright}
{CERN-TH. 6827/93} \\
{PUPT-1382 , LBL-33767 , UCB-PTH-93/09, hep-th/ 9303046}
 \end{flushright}
%\vskip .3truein
\begin{center}
{{\Large    \bf Twists and Wilson Loops in  the String Theory of Two
Dimensional QCD }}
\footnote{This work was supported in part by the Director, Office of
Energy
Research, Office of High Energy and Nuclear Physics, Division of High
Energy Physics of the U.S. Department of Energy under Contract
DE-AC03-76SF00098 and in part by the National Science Foundation
under
grant PHY90-21984.}
\vskip .1in
David J. Gross

{\small {\em
\begin{tabular}{ccc}
\makebox[3in]{Joseph Henry Laboratories, Princeton University } & &Theory
Division\\
Princeton, New Jersey 08544,  & and & \makebox[2.7in]{ CERN, CH-1211} \\
  gross@puhep1.princeton.edu& & Geneva 23, Switzerland \\
\end{tabular}}}

\vskip .1in
Washington Taylor IV
{\small {\em
\begin{tabular}{ccc}
\makebox[ 3in]{Department of Physics, University of California } & &
\makebox[ 2.7in]{Theoretical
Physics Group}\\
Berkeley, California 94720 & and & Lawrence Berkeley Laboratory,\\
 wati@physics.berkeley.edu & &  Berkeley, California 94720   \\
\end{tabular}}}
\vskip .2 truein

{\bf ABSTRACT}
\end{center}
The string theory that describes two-dimensional QCD in an asymptotic
$1/N$ expansion is investigated further.  A complete geometrical
description of the QCD partition function on an arbitrary manifold is
given in terms of maps of a two dimensional orientable surface onto
the target space. This includes correction terms that arise on
surfaces with genus $G \neq 1$, that are described geometrically by
the insertion of extra ``twist'' points in the covering maps.  In
addition the formalism is derived for calculating the vacuum
expectation value of an arbitrary product of Wilson loops on an
arbitrary two dimensional manifold in terms of maps of an open string
world sheet onto the target space.

\begin{flushleft}
{CERN-TH. 6827/93} \\
{PUPT-1382 , LBL-33767 , UCB-PTH-93/09}\\
{March 1993}
\end{flushleft}
\newpage
\renewcommand{\thepage}{\arabic{page}}
\setcounter{page}{1}

\section {Introduction}
\setcounter{equation}{0}
\baselineskip 18.5pt

It has been known for some time that the two-dimensional quantum
gauge theory is
exactly soluble  \cite{migdal} and on a manifold ${\cal M} $ of genus $G$ and
area $A$ has a partition function given by \cite{rusakov}
\begin{eqnarray}
{\cal Z}_{\cal M} & = &  \int[{\cal D} A^\mu]
e^{- {1\over 4 {\tilde g}^2} \int_{\cal M} d^2 x\sqrt{g}\
 Tr F^{\mu \nu}  F_{\mu \nu}} \\
 & = &  {\cal Z}  (G,\lambda A, N)  = \sum_{R}^{} (\dim R)^{2 - 2G}
  e^{-\frac{\lambda A}{2 N}C_2(R)}, \label{eq:partition}
\end{eqnarray}
where the sum is taken over all irreducible representations of the
gauge group, with $\dim R$ and $C_2(R)$ being the dimension and
quadratic Casimir of the representation $R$. ($\lambda$ is related to
the gauge coupling $\tilde{g}$ by $\lambda = \tilde{g}^2 N$.)
Throughout this paper, we will assume that the manifold ${\cal M} $
is
endowed with a metric; however, the theory is invariant under
area-preserving diffeomorphisms.

It was conjectured by one of us (DG) that the gauge theory with gauge
group $SU(N)$ is equivalent to a string theory when one carries out
an
asymptotic expansion of (\ref{eq:partition}) in powers of $1/N$.  The
string coupling is identified with $1/N$ and the string tension is
given by $\lambda/2$ \cite{gross}.  The strategy adopted to verify
this conjecture was to explore whether the $1/N$ expansion of ${\cal
Z}_{\cal M} $ could calculated by summing over maps of a
two-dimensional world sheet onto ${\cal M}$. The string theory would
then be specified by the types of maps that would emerge and their
weights in the sum.  In a previous paper, we proved that this
conjecture was correct \cite{previous} and identified the maps that
reproduce the QCD partition function.  When the target space is a
torus, we were able to completely describe the partition function in
terms of such a sum over maps; when the genus of the target space is
other than one, there were extra terms in the partition function for
which we had no geometric interpretation.
In this paper we complete the discussion of the partition
function on manifolds of arbitrary genus and extend the string theory
to deal with Wilson loops.

Let us recall the results of \cite{previous}.  Define $\Sigma_G$ to
be
the set of all continuous maps $\nu$ that map a (possibly
disconnected) Riemann surface onto ${\cal M} $ such that the map
$\nu$
is locally a covering map at all but a finite set of points of ${\cal
M} $, with $n$ ($\tilde n$) sheets whose orientation is the same as
(opposite of) the orientation of ${\cal M} $.  The asymptotic
expansion of the partition function of the $SU(N)$ gauge theory on
${\cal M} $, in powers of $1/N$, is then given by
\begin{equation}
{\cal Z}(G, \lambda A, N)= \sum_{\nu \in \Sigma_G}
\frac{(-1)^{\tilde t}}{| S_{\nu}|}
e^{- \frac{ (n + \tilde{n}) \lambda A}{2} }
\frac{(\lambda A)^{(i+ t +\tilde t+ h)}}{i!t!\tilde t ! h!}
N^{(n + \tilde{n})(2- 2G)- 2(t +\tilde t +  h) - i} \left[  1
+ {\cal O}\left(  {1\over N } \right) \right],
\label{eq:total}
\end{equation}
where $| S_\nu |$ is the order of symmetry of the map $\nu$ and $i,t,
\tilde{t}, h$ are the quantities of each of the possible types of
point singularities that can appear in the map.  These singularities
include branch points ($i)$, that can lie anywhere on ${\cal M} $;
contracted tubes that connect sheets of the cover with the same ($t$)
and opposite ($\tilde t$) orientations and contracted handles ($h$)
of
the covering space.  The tubes and handles are mapped into points of
${\cal M} $ and are thus infinitesimal.  The factor of $(\lambda
A)^{(i + t + \tilde{t}+ h)}/i!t!\tilde{t}!h!$ corresponds to the
range
of possible locations of the singularity points in ${\cal M} $.  The
power of $1/N$ is precisely $2g-2$, where $g=1 + (n + \tilde{n})(G-1)
+t +\tilde t + h +i/2$\  is the genus of the covering space (the
world
sheet of the string).  The ${\cal O}(\frac{1}{N} ) $ corrections in
this formula only occur when the genus of ${\cal M} $ is other than
one.

An important feature of these maps is that they do not contain folds
of non-zero length.  This is an enormous simplification of the set of
continuous maps between two-dimensional manifolds, and is the stringy
reason, we believe, for the simplicity of QCD${}_2$.  The suppression
of folds means that all the maps in (\ref{eq:total}) have winding
number greater than zero, since maps with winding number zero
necessarily contain folds. This explains why our string theory,
unlike
any other, does not contain any propagating particles.  It is clear
that space-time histories that describe particle propagation
correspond to maps with winding number zero. We must forbid such maps
if we want to describe pure QCD${}_2$ which has no propagating
states.
Thus in our string theory even the center of mass degree of freedom,
the so-called ``tachyon'', is absent.  This bodes well for the analog
string theory of QCD in higher dimensions, which remains to be
constructed.  Once the center of mass degree of freedom is absent in
two dimensions there is no reason to expect a tachyon to appear in
the
higher dimensional theory. Presumably the dynamics of the string,
which in two dimensions prevents folds that have infinite extrinsic
curvature, will suppress extrinsic curvature in higher dimensions and
eliminate the usual instabilities. Perhaps for the same reason
graviton and dilaton modes will not appear. There is certainly no
indication of them in our two dimensional theory. That our string
theory does not contain the global gravitational and dilaton degrees
of freedom found in Liouville string theories in two dimensions is
indicated by the fact that it exists for any target space manifold
and
is insensitive to the metric of the target space.  These features
certainly befit a QCD string theory, which should not be a theory of
gravity and should not contain either a dilaton or a tachyon.

Another feature of the string theory we are investigating is the
existence of two separate chiral sectors, corresponding to covering
spaces of two opposite orientations, as is appropriate for covering
maps of a two dimensional manifold by a orientable world sheet. These
sectors correspond in QCD to the contribution of quark-like
representations of $SU(N)$ and their complex conjugates,
anti-quark-like representations.  These two sectors are coupled only
by the infinitesimal tubes connecting sheets of opposite orientation.

The case when (\ref{eq:total}) is exact, that is when the target
space
is a torus, is the case of most physical interest; however, it is also
useful to consider the theory for arbitrary genus.  For genus $G \neq
1$ there are extra terms (and extra maps) that arise from the $1/N$
corrections in the expansion of $ (\dim R)^{2 - 2G}$ in
(\ref{eq:partition}).  In our previous work, these correction terms
were not given a geometrical interpretation.  In this paper, we
describe the maps and extra factors that arise when the correction
terms are included.  We show that the extra terms can be described
geometrically by the insertion of $|2-2G|$ extra ``twist'' points in
the covering maps.  At these points, the sheets of each orientation
are permuted in an arbitrary fashion by a singularity equivalent to a
multiple branch point.  Sheets of opposite orientation can also be
connected at twist points by orientation-reversing tubes.  When the
sign of $ I =2-2 G$ is positive, we refer to the twist points as
``$\Omega$-points''.  The weights associated with this type of twist
point are fairly simple.  When the sign of $I$ is negative, however,
the weights associated with the twist points are more complex; we
refer to the twist points in this case as ``$\Omega^{-1}$-points'',
since the effect of one of these twist points exactly cancels the
effect of a single $\Omega$-point.  We do not yet have an complete
understanding of the significance of these points from the string
theory perspective; however, it is hoped that their structure will
give a useful clue towards discovering an action formalism for this
string theory.

One of the other main results of this paper is the complete
formulation of a method for calculating the vacuum expectation value
of an arbitrary product of Wilson loops in a single chiral sector of
the theory.  In order to understand how the expectation value of a
Wilson loop can be described as a sum over covering maps, it is
necessary to define a set of observables for Wilson loops that is
appropriate for the string theory.  In the gauge theory picture, it
is
natural to associate each Wilson loop with a representation $R$ of
the
gauge group.  The Wilson loop around a closed curve $\gamma$ is
defined by
\begin{equation}
W(R,\gamma) = {\rm Tr}_R {\cal P} \left\{ \exp\left(  i\oint_\gamma{
{
\bf A_\mu  }dx^\mu} \right) \right\}\equiv\chi_R\left( U_\gamma
\right),
%\label{eq:}
\end{equation}
where the trace of the path-ordered exponential is taken in the
representation $R$.  In the string theory picture, we would like to
associate the $1/N$ expansion of expectation values of Wilson loops
to
the genus expansion of an open string theory, i.e. to maps of a two
dimensional manifold with boundaries onto the target space on which
there are certain specified curves (the closed paths that define the
Wilson loops). These maps will map the boundaries of the string world
sheet (namely certain circles $S^1$) onto the Wilson loops. Since
our
basic maps are covering maps we would expect that as we go around the
Wilson loop $\gamma$ we will induce some permutation of the sheets
covering the loop.  We will therefore associate with each Wilson loop
an integer $k$, and an element $\sigma$ of the permutation group
$S_k$
on $k$ elements.  Intuitively, $k$ will correspond to the number of
sheets of the world sheet boundary covering the Wilson loop, and $S_k$
will be the permutation on
those sheets that is realized by moving around the loop once.  We
thus want to define a different basis for the Wilson loops, labelled
not by representations of $SU(N)$, but rather by elements of
the symmetric group. If $\sigma$ is a permutation with cycles of size
$k_1, \ldots, k_s$, the natural guess, whose validity we will
establish, is that the appropriate gauge theory observable  to pick
out covering maps with  boundary conditions described by $\sigma$ is
\begin{equation}
\Upsilon_{\sigma} (U) = \prod_{j = 1}^{s} ({\rm Tr}\; U^{k_j}),
%\label{eq:}
\end{equation}
where the matrices $U$ are in the fundamental representation.  The $
\Upsilon$ observables are, like the characters, a complete set of
class functions. In fact they are the well-known {\em Schur
functions}
of the eigenvalues of $U$ and are related to the characters by the
Frobenius relations,
\begin{eqnarray}
\chi_R (U) & = & \sum_{\sigma\in S_k}\frac{\chi_{R}(\sigma)}{k!}
\Upsilon_{\sigma} (U),  \label{eq:frobenius} \\
\Upsilon_{\sigma}(U) & = &  \sum_{R\in Y_k} \chi_{R} (\sigma)
\chi_{R} (U),
\label{eq:frobenius2}
\end{eqnarray}
where $\chi_R (U)$ is the character of $U$ in the representation $R$
of $SU(N)$ associated with a Young tableau containing $k$ boxes, and
$\chi_R (\sigma)$ is the character of the permutation $\sigma$ in the
representation of $S_k$ associated with the same Young tableau.  The
sum in (\ref{eq:frobenius2}) is taken over all representations $R$
whose Young tableaux are in the set $Y_k$ of Young tableaux with $k$
boxes.  Actually we will need to generalize the $\Upsilon_{\sigma}
(U)$ basis and the Frobenius relations to account for the full set of
relevant representations that contribute in the $1/N$ expansion.  The
calculations relevant to this generalization are carried out in the
Appendix. These relations are crucial in establishing the
mathematical
connection between the group theory of QCD and the geometry of the
string
theory.

The final result of our analysis of Wilson loops is that the vacuum
expectation value of an arbitrary product of Wilson loops in a single
sector of the theory can be described by a sum over covering maps
similar to (\ref{eq:total}), where the boundary of the covering space
is a union of covers of each Wilson loop, such that the boundary
covering a Wilson loop $ \gamma$ associated with a permutation
$\sigma$ with $s$ cycles of length $k_1, \ldots, k_s$ is a disjoint
union of $s$ loops that cover $\gamma$ exactly $k_1, \ldots, k_s$
times.  Each of the regions into which the set of Wilson loops
divides
${\cal M} $ is also associated with a number of $\Omega$-points or
$\Omega^{-1}$-points.  Although the sum over all regions of the
number
of $\Omega$-points must be $2-2 G$ (counting $\Omega^{-1}$-points as
$-1$), these points are distributed according to the structure of the
Wilson loops, so that even when $G =1$ it may be necessary to
incorporate the $1/N$ correction terms due to twist points.  Thus,
even to understand the theory on a torus it will be necessary to
understand how the twists are related to the string theory action.

In the next section, we will review the results from \cite{previous}
and describe how the corrections from twist points enter into the
exact formula for the partition function for arbitrary genus in a
single chiral sector of the theory.  In Section 3 we discuss the
exact
partition function in the complete theory, including both chiral
sectors.  In order to deal with the most general relevant
representations of $SU(N)$, we use the generalized Frobenius
relations
described in Appendix A.  In Section 4, we derive expressions for the
partition function on manifolds with boundaries, and describe how the
partition function behaves when two manifolds are glued along a
common
boundary component.  We use these results in section 5, where we
derive the general formula for the vacuum expectation value of a
product of Wilson loops in a single chiral sector.  In section 5, we
also discuss the problem of calculating Wilson loops in the coupled
theory.  We present an algorithm that in principle produces the VEV
of an arbitrary Wilson loop in the coupled theory, and describe some
specific types of Wilson loops whose complete VEV's can easily be
computed.  In Section 6 we give a number of simple examples of the
$1/N$ expansions of the partition function as constructed by summing
maps.  In Section 7 we give some simple examples of Wilson loop
vacuum
expectation values. Section 8 contains an outline of how one can
include dynamical quarks into the string picture.  Finally we
conclude
with a summary of our results and discuss some further questions.

\section {The partition function}
\setcounter{equation}{0}
\baselineskip 18.5pt

The partition function for QCD$_2$ (\ref{eq:partition}) is expressed
as a sum
over
all representations $R$ of the gauge group.  Since the gauge group
itself changes as we vary $N$, in order to construct an asymptotic
expansion for the partition function in powers of $1/N$, it is
necessary to fix
the
representations so that the dimension and quadratic Casimir
of a representation can themselves be expanded in powers of
$1/N$.  This can be accomplished by the use of Young tableaux to
label the
representations.  Every representation $R$ is associated with a Young
tableau containing some number of boxes $n$, that are distributed in
rows of
length $ {n}_1 \geq {n}_2 \cdots \geq {n}_l > 0$ and columns
of length $c_1 \geq c_2 \cdots \geq c_k>0 $, where $k = {n}_1$, and
$l = c_1$.  We will abuse notation slightly and  sometimes use $R$ to
refer to
the Young
tableau as well as to the representation itself.
The quadratic Casimir of the representation $R$ is given by
\begin{equation}
C_2 (R) = n N + \tilde{C}(R) - \frac{n^2}{N},
\label{eq:casimir}
\end{equation}
where
\begin{equation}
\tilde{C}(R) = \sum_{i} {n}_i^2 - \sum_{i}c_i^2.
%\label{eq:}
\end{equation}
It was shown in \cite{previous} that the term $\tilde{C}(R)$ is
related to a character of the permutation group by
\begin{equation}
\tilde{C}(R) = \frac{n (n-1)\chi_{R} (T_2)}{d_R},
%\label{eq:}
\end{equation}
where $d_R = \chi_R (1)$ is the dimension of the representation of
the
permutation
group $S_n$ associated with the Young tableau $R$, and $\chi_{R}
(T_2)$ is the character in this representation of any permutation
in the conjugacy class $T_2$ of permutations containing a single
cycle of
length 2 and $n -2$ cycles of length 1.

The dimension of the representation $R$ of $SU(N)$ can be expressed
in
a form
appropriate for the large $N$ expansion by setting $U = 1$
in the Frobenius formula (\ref{eq:frobenius}),
\begin{equation}
\dim R = \frac{1}{n!} \sum_{\sigma \in S_n}
N^{K_\sigma} \chi_{R} (\sigma),
\label{eq:dimension}
\end{equation}
where $K_\sigma$ is the number of cycles in the permutation $\sigma$.
Note that the leading order term in this polynomial
arises from the identity permutation ($\sigma= [1^n]$) and is equal
to $d_R
N^n/n!$.
By summing over all $n$,  and for each $n$ summing over
all Young tableaux in $Y_n$, we can write (\ref{eq:partition}) in a
form that
is appropriate  for an asymptotic expansion of the partition function
in powers
of $1/N$, namely
\begin{equation}
Z (G,\lambda A, N)  = \sum_{n= 0}^{\infty}  \sum_{R \in Y_n}
 (\dim R)^{2 - 2G}  e^{-\frac{\lambda A}{2 N}C_2(R)}.
\label{eq:partfunct}
\end{equation}
The idea is to carry out the $1/N$ expansion of $C_2(R)$ and of
$\dim(R)$,
using (\ref{eq:casimir}) and (\ref{eq:dimension}), inside the sum
over
the number
of boxes $n$.

This expansion, however, only contains half of the full theory.  The
representations associated with Young tableaux with a fixed number of
boxes arise from symmetrizing and antisymmetrizing tensor products of
the fundamental representation of $SU(N)$.  It is also necessary to
include representations that arise from tensor products of the
conjugate representation $U \rightarrow U^{\dagger}$;  these
representations also have quadratic Casimirs with leading terms of
order  $N$.  The most general representation that we must consider
is
a composite representation that we get  by taking the largest
irreducible
representation contained in the tensor product of a representation
$R$
with a finite number of boxes, and the conjugate $\bar{S}$ of another
representation $S$, that also has  a finite number of boxes.
Specifically,
given
two representations $R$ and $S$, whose Young tableaux contain $n$ and
$\tilde n$ boxes respectively, with rows
of length $n_i$ and $\tilde n_i$, and columns
of length $c_i$ and $\tilde c_i$,
we can form a
new representation $\bar{S}R$, with column lengths
\begin{equation}
\left\{ \begin{array}{ll}
N - \tilde c_{ \tilde{n}_1 + 1 - i}, &  i \leq  \tilde{n}_1\\
c_{i - \tilde{n}_1}, & i > \tilde{n}_1
\end{array}\right\}.
%\label{eq:}
\end{equation}
The quadratic Casimir of the representation $\bar{S}R$ is given by
\begin{equation}
C_2 (\bar{S}R) = C_2 (R) + C_2 (S)+ \frac{2n\tilde n}{N} .
\label{eq:sumcasimir}
\end{equation}
It can also be shown that the dimension of the representation
$\bar{S}R$ is
given by
\begin{eqnarray}
\dim \bar{S}R & =&  \dim S \dim R \prod_{i,j} { \left(  N+1 -i-j
\right)
\left(   N+1
-i-j +n_i+\tilde n_j \right)    \over  \left(  N+1 -i-j +n_i\right)
\left(  N+1
-i-j +\tilde n_j  \right)}  \\
&  =& \dim S \dim R + {\cal O}  (\frac{1}{N^2}).
\label{eq:compositedimension}
\end{eqnarray}
Using the results of Appendix A we shall present a more useful
formula
for the
$\frac{1}{N^2}$ corrections to $\dim \bar{S}R $ in the next section.

By including all composite representations, which are the only
representations of $SU(N)$ whose quadratic Casimir grows like $N$, we
can write
 the partition function of the full gauge theory as
\begin{equation}
{\cal Z}  (G,\lambda A, N)  =  \sum_{n}\sum_{\tilde n}\sum_{R \in
Y_n}
\sum_{S \in Y_{\tilde n}}^{} (\dim \bar{S}R)^{2 - 2G}
  e^{-\frac{\lambda A}{2 N} \left[  C_2(R)+ C_2 (S)+ \frac{2n \tilde
n}{N}
\right]},
\label{eq:cpartition}
%\label{eq:}
\end{equation}
and derive the asymptotic  expansion by expanding the Casimirs and
dimensions
in powers of $1/N$ inside the sum. The
%over $n$ and $\tilde n$
corrections to
this expansion are then of order $\exp[-N]$, corresponding to
non-perturbative
string
corrections of order $\exp[-1/g_{\rm string}]$, that are invisible
in string
perturbation theory.

This partition function can be factored into two copies of the
partition function (\ref{eq:partfunct}), that are only coupled by
the term $\frac{2n \tilde{n}}{N} $ in the quadratic Casimir and by
some terms of order $1/N^2$ that arise from the non-leading terms in
the
expansion of the dimension.  We refer to the two
factors in this product as chiral sectors of the theory.  We
interpret the two
chiral sectors as corresponding to orientation-preserving and to
orientation-reversing string maps;  under this interpretation,
the coupling terms in  the
exponential correspond to infinitesimal orientation-reversing tubes
that
connect covering sheets
of opposite relative orientation.

To understand how
(\ref{eq:cpartition}) can be understood in terms of a string theory,
it is simplest to analyze a single sector of the theory first, and
then to describe how the sectors are coupled.  Therefore, let us
now consider a single chiral sector of the theory, with partition
function (\ref{eq:partfunct}).  Breaking up the term $n^2/N$ from the
quadratic Casimir into a sum of terms $n/N + n (n -1)/N$
\cite{Minahan}
, and
expanding the exponential of the quadratic Casimir, this partition
function, which we denote by $Z^+$,  can be rewritten as
\begin{eqnarray}
Z^+ (G,\lambda A, N)& = & \sum_{n = 0}^{\infty}  \sum_{R \in Y_n}
e^{- \frac{n \lambda A}{2}}\sum_{i,t,h}
\frac{(\lambda A)^{i + t + h}}{i!  \;t!  \; h!}
\frac{(-1)^i n^h (n^{2} - n)^{t+ i} }{(n!)^{2-2G}\; 2^{i + t + h}\;
d_R^{i}}
\left(\chi_R (T_2) \right)^i   \nonumber\\
 & &\hspace{1.3in} \cdot N^{n (2- 2G)- i -2 (t + h)}
\left(\sum_{\sigma \in S_n}
\frac{\chi_{R} (\sigma)}{N^{n - K_{\sigma}}}  \right)^{2-2 G}.
\label{eq:expansion}
\end{eqnarray}

In order to rewrite (\ref{eq:expansion}) in terms of a sum over maps,
it will be useful to recall some elementary results from the
representation theory of the symmetric group $S_n$.  One result we
will use repeatedly is that when $T$ is a conjugacy class of elements
in $S_n$ (i.e. $ \sigma' \in T_\sigma$ if $\sigma =s \sigma s^{-1}$
for $s \in S_n$),
and $\rho$ is an arbitrary element of $S_n$, the characters
of $\ T$ and $\rho$ in a fixed representation $R$ can be combined
according to the formula
\begin{equation}
\sum_{\sigma \in T}\frac{\chi_{R} (\sigma)}{d_R}  \chi_{R} (\rho) =
\sum_{\sigma \in T} \chi_{R} (\sigma \rho).
\label{eq:identity1}
\end{equation}
This result follows immediately from the fact that in any
representation $R$ the sum of the matrices associated with the
elements in $T$ must be proportional to the identity matrix,  since
this sum commutes with all elements of $S_n$

	Associated with the symmetric group, as with any finite
group, there
is a natural group algebra, consisting of all linear combinations of
group elements with real coefficients.  Elements of the group algebra
are added and multiplied in the natural way.    A Kronecker-type
delta function
can be defined on
the symmetric group algebra; if $\sum_{\sigma}c_\sigma \sigma$ is an
element of the group algebra, then we define the delta function by
\begin{equation}
\delta (\sum_{\sigma}c_\sigma \sigma) = c_{1},
%%\label{eq:}
\end{equation}
where $c_{1}$ is the coefficient of the identity
permutation.

Other results from the representation theory of $S_n$ that we will
need are
\begin{equation}
\delta (\rho) = \frac{1}{n!} \sum_{R}^{} d_R \chi_R (\rho),
\label{eq:identity2}
\end{equation}
which follows from the completeness of the characters, and the
relation
\begin{equation}
(\frac{n!}{d_R} )^{2}= \sum_{s,t \in S_n}
\frac{\chi_{R} (sts^{-1} t^{-1})}{d_R},
\label{eq:identity3}
\end{equation}
that may be proven using (\ref{eq:identity1}).

Finally there is an element
$\Omega_n$ of
the group algebra on $S_n$, defined by (recall that $K_\sigma=$
number
of cycles of $\sigma$),
\begin{equation}
\Omega_n = \sum_{\sigma \in S_n} \sigma N^{K_\sigma- n},
\label{eq:omega}
\end{equation}
that will be useful in understanding the effects of the dimension
term in the partition function.
In fact, the formula  for the dimension (\ref{eq:dimension}) can be
written as
\begin{equation}
 \dim R = {N^n\over n!} \chi_R(\Omega_n).
\label{eq:dimtwo}
\end{equation}
Since we will also encounter inverse powers of the dimension it is
useful to
invert $\Omega_n$.
Since $\Omega_n$ is a polynomial in
$1/N$, with a leading term of 1 times the identity permutation (in
fact
$\delta (\Omega_n)=1$),
we can write
\begin{equation}
\Omega_n = 1+ \tilde{\Omega}_n,
%\label{eq:}
\end{equation}
where $\tilde{\Omega}_n$ is of order $1/N$.
$\Omega_n$ has a formal inverse $\Omega_n^{-1}$.
$\Omega_n^{-1}$ is an infinite
series in $1/N$, that also has a leading term of 1 times the
identity
permutation,
\begin{eqnarray}
\Omega_n^{-1}   & = & \sum_{\sigma}\omega_\sigma \sigma = \sum_{k =
0}^{\infty}  (-1)^k \tilde{\Omega}_n^k  \label{eq:ominverse}\\
& = &{  \bf 1} \left( 1+O({1\over N^2})\right) -{1\over
N}\sum_{\sigma \in T_2} {\bf \sigma} \left(1+ O({1\over N})\right)
+\cdots
\end{eqnarray}
Since in any representation $R$, $\Omega_n$ is proportional to the
identity matrix,
it follows that
\begin{equation}
\frac{1}{ \chi_R(\Omega_n)} =
\frac{1}{d_R^2} \chi_R(\Omega_n^{-1}).
%\label{eq:}
\end{equation}
Thus,
we can write
\begin{equation}
 {1 \over \dim R} = { n!\over N^n d_R^2} \chi_R(\Omega_n^{-1}).
\label{eq:dimthree}
\end{equation}
More generally,
\begin{equation}
 ({ \dim R})^m = \left({  N^n d_R\over n!  }  \right)^m {
\chi_R(\Omega_n^{m})\over d_R} \  ,
\label{eq:dimfour}
\end{equation}
for both positive and negative $m$.

{}From equations (\ref{eq:identity1}), (\ref{eq:identity2}),
(\ref{eq:identity3}), and (\ref{eq:dimtwo}), the asymptotic expansion
(\ref{eq:expansion}) can be
rewritten in the form
\begin{eqnarray}
Z ^+(G,\lambda A, N)& = &  \sum_{n, i, t,h}
e^{- \frac{n \lambda A}{2} }
\frac{(\lambda A)^{i + t + h}}{i!  \;t!  \; h!}
N^{n (2 - 2G)- i-2 (t + h)}
\frac{(-1)^i n^h (n^{2} - n)^t}{2^{t + h}}
\nonumber \\
& & \cdot \sum_{p_1, \ldots ,p_i \in T_2}
\sum_{s_1, t_1, \ldots ,s_G,t_G\in S_n} \left[
\frac{1}{n!}   \delta (p_1\cdots p_i \Omega_n^{2 - 2G}
\prod_{j =1}^{G}s_j t_j s_j^{-1} t_j^{-1}
)\right].
\label{eq:result}
\end{eqnarray}
This equation is derived from (\ref{eq:expansion}) by
applying (\ref{eq:identity3}) $G$
times, repeatedly applying (\ref{eq:identity1}) until all characters
are
combined into a single character, and then applying
(\ref{eq:identity2}) to turn the sum over
representations into a single delta function.

It is now possible to give a geometric interpretation to the
partition
function by showing that (\ref{eq:result}) is equal to a sum over
covering maps of ${\cal M} $.  The most unusual part of this
interpretation is the effect of the $\Omega_n$ terms in the delta
function.  These terms correspond to having extra twists in the
covering at precisely $ |I| =|2 -2 G |$ points in ${\cal M} $.  To be
more specific, let us choose an arbitrary set of points $z_1, \ldots,
z_{|I|}$ in ${\cal M} $.  For a fixed value of $i$, we also select a
set
of branch points $w_1, \ldots, w_i\in {\cal M} $.    For the moment,
let us assume that $G \leq 1$ so $ I$ is positive.  In this case,
we assert that for a fixed value of $n$, the
expression
\begin{equation}
 \sum_{p_1, \ldots ,p_i \in T_2}
\sum_{s_1, t_1, \ldots ,s_G,t_G\in S_n}
N^{n(2-2 G)- i}
\left[\frac{1}{n!}   \delta (p_1\cdots p_i \Omega_n^{2 - 2G}
\prod_{j}s_j t_j s_j^{-1} t_j^{-1}
)\right]
\label{eq:case}
\end{equation}
is precisely equal to the sum of a factor $N^{2-2 g}/| S_\nu |$ over
all $n$-fold covers $\nu$ of ${\cal M} $ that have single branch
points at each of the points $w_1, \ldots, w_i$, and that have
arbitrary twists (permutations of sheets) arising from multiple
branch
points at $z_1, \ldots, z_{|I|}$, where $g$ is the genus of the
covering
space, and $| S_\nu |$ is the symmetry factor of the cover (the
number
of homeomorphisms $\pi$ of the covering space that satisfy $\nu \pi
=
\nu$).

This assertion can be proven by considering the fundamental
group  of the manifold ${\cal M}$ with punctures at the points
$z_j,w_j$,
 $ {\cal G} =\pi_1 ({\cal M}\setminus\{z_1, \ldots, z_{|I|},w_1,
\ldots, w_i
\})$.  This group is generated by a set of loops $a_1, b_1, \ldots,
a_G,b_G$ that are a basis for the homology of ${\cal M} $, a set of
loops $c_1, \ldots, c_i$ around the branch points, and a set of loops
$d_1, \ldots, d_{| I |}$ around the twist points.  The only relation
satisfied by these generators is
\begin{equation}
c_1 \cdots c_i d_1\cdots d_{| I |}
a_1 b_1 a_1^{-1} b_1^{-1} a_2 b_2 a_2^{-1} b_2^{-1} \cdots
a_G b_G a_G^{-1} b_G^{-1}=1.
%\label{eq:}
\end{equation}
Assign  a set of labels $1, \ldots, n$ to the sheets of a
covering space at a fixed basepoint. As one   transverses the
manifold around
any loop that is not homotopically trivial the labeling of the sheets
is permuted. We will define the permutations
on the covering sheet  labels that
are realized by the loops $a_j,b_j,c_j,d_j$ to be elements of the
permutation group  of $n$ objects denoted by
$s_j,t_j,p_j,q_j$
respectively. We
see then that each cover corresponds to a homomorphism $H$ from
${\cal G}$ into the symmetric group $S_n$.
Note that the branch points must be associated with permutations in
the conjugacy class $T_2$, since they are assumed to be branch points
with branching number 1.
Any two homomorphisms $H$
and $H'$ that are related through conjugation by an element of the
symmetric group, $H' = \rho H \rho^{-1}$, correspond to distinct
labelings of a single covering space.  Every symmetry of the covering
space corresponds to a permutation that leaves $H$ invariant.  Since
there are exactly $n!$  possible labelings of a given cover, in order
to sum over all covers a factor of $1/| S_\nu |$, it will suffice to
sum over all homomorphisms $H$ a factor of $1/n!$.  Writing the
sum of $N^{2-2 g}/n!$
over all homomorphisms gives exactly the sum in (\ref{eq:case}),
where
each  branch point contributes -1 to the  Euler characteristic $2-2
g$
of the covering
space, and where each twist $q_j$ contributes $K_{q_j}- n$.  For
a more detailed calculation of this type, see \cite{previous}.

We can thus  rewrite the partition function (\ref{eq:result})
completely in terms of a sum over covers when $G \leq 1$.  We define
the set of orientation-preserving
covers of ${\cal M} $ with $|I |$ twist points to be
$\Sigma_+ ({\cal M} )$.  We include in this set all maps from a
Riemann surface ${\cal M}_g $ of genus $g$ onto ${\cal M} $ that are
orientation-preserving maps, and that are covering maps in a local
neighborhood of all points in ${\cal M} $ except for a finite number
of singular points.  The singular points are divided into $ |I|$
twist
points,  $i$ branch points, $t$ points where loops in ${\cal M}_g $
are mapped to single points in ${\cal M} $ (orientation-preserving
tubes), and $h$ points where entire handles of ${\cal M}_g $ are
mapped to single points in ${\cal M} $.
The twist points are fixed points $z_j$ in ${\cal M} $ where
the covering map takes a set of neighborhoods in ${\cal M}_g $ into a
single
neighborhood of $z_j$, such that the preimage of $z_j$ in each of the
neighborhoods in ${\cal M}_g $ is a single point.  These points can
be
viewed as points where multiple single branch points have combined
to give an arbitrary permutation of the sheets of the covering space,
without adding handles or tubes.
%however not all combinations of single branch points are allowed; a
%given twist $q_j$ must be produced by the composition of exactly $n
%K_{q_j}$ branch points.

The set $\Sigma_+({\cal M})$ can be separated into disjoint
components labeled by the integers $i,t,h$.  Each of these
components breaks down further into a set of disjoint connected
components labeled by topological type, each of which
is a continuous manifold parameterized by the positions of
the branch points $w_j$, and the positions of the tubes and handles,
$x_j$ and $y_j$.  On each connected component of  $\Sigma_+({\cal
M})$,
we can define an integration measure $d \nu$ for the parameters
$w_j,x_j,$ and
$y_j$  that is proportional to a product of the area measures for
each of the parameters.  Since the branch
points, tubes, and handles are indistinguishable, by integrating over
the parameters we get a factor of $(\lambda A)^{(i + t +
h)}/i!t!h!$  when the proportionality constant is $\lambda$.
We define an integral over all of $\Sigma_+ ({\cal M} )$ to be a sum
over all connected components of $\Sigma_+ ({\cal M} )$ of the
integral
with measure $d \nu$ over each connected component.  In terms of such
an integral, the partition function
(\ref{eq:result}) is simply given by
\begin{equation}
Z ^+(G,\lambda A, N) = \int_{\Sigma_+ ({\cal M})}  d \nu
\; e^{- \frac{n \lambda A}{2} }
\frac{(-1)^{i} N^{2 -2 g}}{ |S_\nu |} ,
\label{eq:partitionsimple}
\end{equation}
where $n$ is the winding number of the map $ \nu$,
 $i$ is the number of branch points in $\nu$, $| S_\nu |$ is
the symmetry factor, and $g$ is the genus of the covering space.  The
extra factors of $n/2$ and $n (n - 1)/2$ in
(\ref{eq:result}) appear because a handle can be mapped onto any of
the $n$ sheets of the cover and has an extra symmetry factor of 2,
and
a tube can connect an arbitrary pair of the sheets.

We have now given a simple formula for the partition
function  on a manifold with genus $G \leq 1$ in
terms of a sum over covering maps.  It is not hard to see how this
formula must be modified if
the genus is larger than $1$.  When $I$ is positive, the algebra
element $\Omega_n$ appears in the expression for the partition
function to a positive power.  Each instance of $\Omega_n$ in the
delta function in (\ref{eq:result}) corresponds to a twist point in
${\cal M} $ where all twists $q_j$ are given a weight of
$N^{K_{q_j}- n}$.
We will refer to points of this type as ``$\Omega$-points''.  For
instance, to calculate the partition function on a sphere, one must
sum over all covers with two $\Omega$-points.  When $I$ is negative,
the number of factors of $\Omega_n$ in the delta function is
negative.
This can be interpreted as indicating that we must sum over covers
that contain twist points with a different set of weights.
We will refer to the $ -I$ twist points of this new type as
``$\Omega^{-1}$-points''.   If
we write the coefficients of the algebra element $\Omega_n^{-1}$ as
$\omega_\sigma$,  as in (\ref{eq:ominverse}),
then the weight of  an $\Omega^{-1}$-point with a permutation $q_j$
will be
$\omega_{q_j}$.  The coefficients $\omega_{\sigma}$
are
infinite series in $1/N$.
A simpler way to interpret these $\Omega^{-1}$-points is to use the
expression  (\ref{eq:ominverse}) for $\Omega_n^{-1}$ in terms of
$\tilde{\Omega}_n$.  From this expression, it is clear that an
$\Omega^{-1}$-point can be viewed as a combination of an arbitrary
number ($x\geq 0$) of nontrivial twists such as would arise at an
$\Omega$-point, with each twist carrying a factor of $-1$.
An $\Omega^{-1}$-point is then a point where an arbitrary number of
nontrivial twists coalesce to form a singularity with an arbitrary
permutation on the sheets, and possibly internal handles or tubes.
By using the same logic as in the case $G \leq 1$, the asymptotic
expansion for a single chiral sector of the gauge
theory on a manifold with an arbitrary genus $G$ can be written as
\begin{equation}
Z^+(G,\lambda A, N) =\int_{\Sigma_+ ({\cal M} )} d \nu
\;  e^{- \frac{n \lambda A}{2} }
\frac{(-1)^{i} N^{2 -2 g}}{ |S_\nu |}  \prod_{j=1 }^{\max (0, - I)}
(-1)^{x_j},
\label{eq:general}
\end{equation}
where $x_j$ are the numbers of nontrivial twists at the
$\max (0, - I) $
$\Omega^{-1}$-points in the map $\nu$, and $n,i,|S_\nu |$ and $g$ are
as above.

Although the introduction of twist points
($\Omega$-points and $\Omega^{-1}$-points) gives us a geometric
interpretation of all terms in the partition function for the theory
on an arbitrary genus Riemann surface, it is rather unclear how these
points might be described by a string theory action.  Since locations
of the twist points are not  integrated over (the twist points do not
carry factors of the area), they seem to correspond to some global,
rather than local structure.  In fact, although we have described the
twists as occurring at fixed points in the manifold ${\cal M} $, they
may in fact describe some purely non-local structure.
Because the number of $\Omega$-points is equal to the Euler
characteristic of the manifold ${\cal M} $, it is tempting to
speculate that these twist points are somehow related to the divisor
of the tangent bundle of ${\cal M} $.
One might hope to find
a suitable way of coupling fermions to the theory in a way that
would
simultaneously cancel folds and give rise to terms like those coming
{}from $\Omega$-points and $\Omega^{-1}$-points; however, we have not
yet
succeeded in finding a string action that has  all these desired
properties.

\section{The Complete Theory}
\setcounter{equation}{0}
\baselineskip 18.5pt

The analysis performed in the previous section applied only to a
single chiral sector of the theory.  It is possible to combine the two
chiral sectors, and to use a similar formalism in describing the
partition function for the complete theory.  When we combine the two
sectors, we can consider the theory to include independent sums over
all covering maps that are orientation-preserving and
orientation-reversing.  These sets of covering maps are coupled. The
simplest coupling arises from the extra term $2n \tilde{n}/N$ in the
quadratic Casimir (\ref{eq:sumcasimir}). It has a natural
interpretation as corresponding to an arbitrary number of
orientation-reversing tubes connecting sheets of covers from the two
sectors. These tubes are mapped onto single points of the target space
${\cal M}$.  The positions of these tubes are integrated over just as
are the positions of orientation-preserving tubes in the two sectors.
Thus each tube contributes a factor of $ n \tilde n \lambda A/N^2$,
since the genus of the map is increased by one for each additional
tube.  In addition it appears that we must include a minus sign each
time such a tube occurs. The contribution of the orientation reversing
tubes then exponentiates to yield a factor of $\exp[ - (\lambda A/ 2)
\; 2n \tilde{n}/N^2]$ in the partition function, reproducing the coupling
term in the exponent in (\ref{eq:cpartition}).

There are also extra terms in the formula for the dimension of a
composite representation (\ref{eq:compositedimension}) that we argue
correspond to additional tubes connecting the two sectors. To derive
these we need to determine the $1/N$ expansion of the dimension of the
composite representation $\bar S R$. This analysis is carried out in
Appendix A, where we generalize the Frobenius relations for the
characters of composite representations of $SU(N)$ and derive the
$1/N$ expansion of the dimensions of these representations. The
formula (\ref{eq:compdim}) derived there can be written as
\begin{equation}
\dim\bar S R = {1\over n! \tilde n! }\sum_{\sigma, \tau}N^{K_\sigma
+K_\tau}
\chi_R(\sigma) \chi_R(\tau) \left(
\sum_{t,j_u,\tilde{j}_u}\prod_{u=1}^{t}(-\frac{k_{j_u}}{N^2})
\right),
\label{eq:compdimtwo}
\end{equation}
where $k_j, \tilde{k}_j$ are the lengths of the cycles of $\sigma,
\tau$ respectively. The sum is taken over all disjoint pairs
$(j_u, \tilde{j}_u)$ of cycles of
$\sigma$ and $\tau$  that are of the same length. This means that we
sum over
all $j_u$ and $\tilde j_u$
such that  for $u \neq u'$, $j_u \neq j_{u'}$
and $\tilde{j}_u \neq \tilde{j}_{u'}$, and such that for all $u$,
$k_{j_u}= \tilde{k}_{\tilde{j}_u}$.

Using this formula we can go through the same steps as
before to
write the complete partition function as a sum over maps.
We define a more general object $\Omega_{n \tilde{n}}$ to be an
element of the algebra of the tensor product group
$S_n \otimes
S_{\tilde{n}}$, given by
\begin{equation}
\Omega_{n \tilde{n}} = \sum_{\sigma, \tau}
\sigma \otimes \tau
\sum_{\upsilon, \upsilon'}
(-1)^{K_{\upsilon}} C_\upsilon
N^{K_{\sigma \setminus \upsilon}+K_{\tau \setminus \upsilon} - n -
\tilde{n}}.
\label{eq:generalomega}
\end{equation}
Similar to the case of a single sector, we have
\begin{equation}
\dim \bar{S}R =  \chi_{R  \otimes S} (\Omega_{n \tilde{n}}).
%\label{eq:}
\end{equation}
Again, $\Omega_{n \tilde{n}}$ is a polynomial in $1/N$ with leading
term $1$, and can be written
\begin{equation}
\Omega_{n \tilde{n}} = 1+ \tilde{\Omega}_{n \tilde{n}}.
%\label{eq:}
\end{equation}
We can expand $\Omega_{n \tilde{n}}^{-1}$ as a power series in $1/N$,
and write
\begin{equation}
\Omega_{n \tilde{n}}^{-1} = \sum_{x = 0}^{\infty}  (-1)^x
\tilde{\Omega}_{n \tilde{n}}.
%\label{eq:}
\end{equation}
The partition function in the complete theory on an arbitrary genus
surface can now be written as a sum over maps,
\begin{equation}
{\cal Z} (G,\lambda A, N) =\int_{\Sigma ({\cal M} )} d \nu
\;  e^{- \frac{(n + \tilde{n}) \lambda A}{2} }
\frac{(-1)^{i + \tilde{t}} N^{2 -2 g}}{ |S_\nu |}
\prod_{j=1 }^{\max (0, - I)}
(-1)^{x_j},
\label{eq:generalcomplete}
\end{equation}
where $\Sigma ({\cal M} )$ is the set of all covering spaces of
${\cal
M}$ with a finite number of singular points, $\tilde{t}$
orientation-reversing tubes, and $2 -2G$
$\Omega$-points.  In the coupled theory an $\Omega$-point is a
point at which in each sector the sheets can be twisted arbitrarily,
and where there can also be orientation-reversing tubes connecting
cycles of these twists, subject to the conditions that no cycle can
be
attached to more than one tube, and tubes can only connect cycles of
the same order.  (These conditions on the placement of tubes follow
{}from (\ref{eq:generalomega}).)    As usual, these
orientation-reversing tubes carry a factor of $-1$.
In the coupled theory,
$\Omega^{-1}$-points again may consist of an arbitrary number of
nontrivial twists, each carrying a factor of $-1$, where among
nontrivial twists we include all possible singularities that might
occur at an $\Omega$-point, including the case where the permutations
in both sectors are trivial but there is at least one tube connecting
sheets of opposite orientation (cycles of length one).

This completes the derivation of the string theory representation,
written as a
sum over maps, for the two-dimensional QCD partition function, to all
orders in
$1/N$ and on an arbitrary space-time background. Except for the
$\Omega$-points all the features of the maps are very simple---the
maps are
simply minimal area maps with no folds, consisting of  covers of the
target
space with possible branchpoints, contracted handles  and tubes
mapped to a
point. The counting of the maps is the natural topological counting.
On
surfaces of genus $G \neq 1$ there are in addition the extra
maps with
twists and tubes connecting the twist points at  $|2-2G| $ fixed
points on the
manifold.
The challenge is to find a string Lagrangian whose partition function
yields
this set of maps.

%\eject

\section{Gluing manifolds}
\setcounter{equation}{0}
\baselineskip 18.5pt

In this section, we will calculate the partition function of
QCD$_2$ on manifolds with boundaries in terms of a sum over string
maps, and derive a set of formulae for
how these partition functions combine when manifolds are glued
together along a common boundary region.  These results will be used
in the next section to calculate vacuum expectation values of Wilson
loops.  The equations in this section can also be used to describe
the
QCD$_2$ string theory in a more geometric fashion, by gluing together
the fundamental objects described by a sphere with 1,2, and 3
boundary
components (caps, propagators, and vertices).  In this section, as in
the next, we will concentrate on a single sector of the theory;
however, we give generalizations of all results to the coupled theory
with the exception of the gluing rule along an edge segment (which is
not used in the Feynman diagram calculations).
An analogous gluing equation can be derived in the coupled theory;
however, we do not have a simple closed-form expression for the
result,
which leads to rather
complicated structures in Wilson loop calculations in the full
theory.

First, we will need a formula for the partition function on a
(compact) Riemann
surface ${\cal M} $ of genus $G$ and area $A$
with an arbitrary boundary.  Assume that the manifold ${\cal M} $ has
a boundary $\partial{\cal M}$ that is a disjoint union of $b$ copies
of the circle $S^1$.  We denote these components of the boundary by
$E_1, \ldots, E_b$.  To calculate the partition function on such a
manifold we
must specify boundary conditions on the gauge fields. The holonomies
of the
gauge field around the
boundary curves are given by
\begin{equation}
U_j =  {\cal P} \left\{ \exp\left(  i\oint_{E_j}{  {
\bf A_\mu  }dx^\mu} \right) \right\},
%\label{eq:}
\end{equation}
where the path-ordered exponential is again taken in the fundamental
representation.  When the holonomies around the boundary curves
are taken to be fixed values $U_1, \ldots U_b$, the partition
function
can be calculated in a similar fashion to the partition function on a
closed manifold, and
is given by \cite{migdal,witten}
\begin{equation}
{\cal Z}  (G,\lambda A, N; \{U_1, \ldots, U_b\})
= \sum_{R}^{} (\dim R)^{2 - 2G-b}
\left(\prod_{j =1}^{b} \chi_{R} (U_j)\right)
e^{-\frac{\lambda A}{2 N}C_2(R)}.
\label{eq:boundarypartition}
\end{equation}

In a single sector of the theory, we can use the Frobenius
relations (\ref{eq:frobenius}) to express the partition function
(\ref{eq:boundarypartition}) in
terms of the class functions $\Upsilon_\sigma (U)$,
%\eject

\begin{eqnarray}
Z^+ (G, \lambda A, N;\{ U_i\}) & = & \sum_{n}\sum_{R\in Y_n}(\dim
R)^{2
- 2G-b}
e^{-\frac{\lambda A}{2 N}C_2(R)} \prod_{j =1}^{b} \chi_{R} (U_j)  \\
 & = & \sum_{n}\sum_{R\in Y_n}(\dim R)^{2 - 2G-b}
 e^{-\frac{\lambda A}{2 N}C_2(R)} \prod_{j =1}^{b}  \left(
\sum_{\sigma_j}\frac{\chi_{R} (\sigma_j)}{n!}\Upsilon_{\sigma_j}(U_j)
\right).
\label{eq:newpart}
\end{eqnarray}
By an argument essentially identical to that of the previous section,
this expression can be rewritten as a sum/integral over the set
$\Sigma_+ (${\cal M}$)$ of orientation-preserving
covers (with  $2 -2G - b \ \Omega$-points and a world-sheet with
boundary) of ${\cal M} $,
\begin{equation}
Z^+ (G,\lambda A, N;\{ U_i\}) =\int_{\Sigma_+ ({\cal M} )} d \nu
\;  e^{- \frac{n \lambda A}{2} }
\frac{(-1)^{i} N^{2 -2 g - \beta}}{ |S_\nu |}
\left(\prod_{j=1 }^{\max (0, - I)} (-1)^{x_j}\right)
\left( \prod_{k =1}^{ b}
\Upsilon_{\sigma_k} (U_k)\right),
\label{eq:bounded}
\end{equation}
where the measure $d \nu$ is defined as in  section 2,
 $\beta = \sum K_{\sigma_k}$ is the number of boundary components of
the covering
space, and
where $\sigma_k$ is in the conjugacy class of the permutation on
sheets
of the cover described by traversing the boundary component $ E_k$ of
${\cal M} $.

A formula that more precisely captures the geometrical nature of
this
partition function is found by evaluating the expectation value of
$\prod_{i=1}^b \Upsilon_{\sigma_i}(U^{\dagger}_i)$. This then yields
\begin{eqnarray}
Z^+ (G,\lambda A, N;\{ \sigma_i\}) & = &<\prod_{i=1}^b
\Upsilon_{\sigma_i}(U^{\dagger}_i)>\\
& = &
\int_{\Sigma_+ ({\cal M};\{\sigma_i\} )} d \nu
\;  e^{- \frac{n \lambda A}{2} }
\frac{(-1)^{i} N^{2 -2 g - \beta}}{ |S_\nu |}
\left(\prod_{j=1 }^{\max (0, - I)}(-1)^{x_j}\right),
\label{eq:b}
\end{eqnarray}
where the condition on the maps $\nu$ in $\Sigma_+ ({\cal
M};\{\sigma_i\} )$ is that when we transverse the
boundary
component $E_i$ the sheets of the cover undergo a permutation in the
class of
${\sigma_i}$.

It is straightforward to generalize (\ref{eq:bounded}) to the coupled
theory.  By using the generalized Frobenius relations from Appendix
A,
the character of an arbitrary composite representation can be
rewritten as a sum of generalized $\Upsilon$ functions. By an
identical analysis to that performed above, we find that the
expectation value of a product $\prod_{i=1}^b
\Upsilon_{\overline{\sigma}_i\tau_i}(U_i,U^{\dagger}_i)$  is given by
the sum over maps
\begin{eqnarray}
{\cal Z}  (G,\lambda A, N;\{ \sigma_i, \tau_i\}) & = &<\prod_{i=1}^b
\Upsilon_{\overline{\tau}_i\sigma_i}(U_i,U^{\dagger}_i)>\\
& = &\int_{\Sigma ({\cal M};\{\sigma_i, \tau_i\} )} d \nu
\;  e^{- \frac{(n + \tilde{n}) \lambda A}{2} }
\frac{(-1)^{i+ \tilde t} N^{2 -2 g - \beta}}{ |S_\nu |}
\left(\prod_{j=1 }^{\max (0, - I)}(-1)^{x_j}\right),
\end{eqnarray}
where the maps in $\Sigma ({\cal M};\{\sigma_i, \tau_i\} )$ are
those covering maps with $2 -2G - b$ $\Omega$-points satisfying the
condition that the
permutations on the orientation-preserving and orientation-reversing
sheets of the cover of the boundary component $E_i$ are given by
$\sigma_i, \tau_i$ respectively.  From this result, it is clear that
in the coupled theory, the generalized $\Upsilon$ functions are the
natural basis for the Hilbert space in the string theory.

\begin{figure}[tbp]
\centering
\begin{picture}(200,100)(- 100,- 50)
\thicklines
\put( - 40,15){\line(1,0){80}}
\put( - 40,-15){\line(1,0){80}}
\put(- 40,0){\oval(10,30)}
\thinlines
\put(0,0){\makebox(0,0){
\begin{picture}( 20, 40)(- 10,- 20)
\thicklines
\put(0,0){\oval(10,30)[ r]}
\thinlines
\put(0,10){\oval(10,10)[ tl]}
\put(0,-10){\oval(10,10)[ bl]}
\put(- 5,- 6){\line(0,1){4}}
\put(- 5,2){\line(0,1){4}}
\end{picture}
}}
\put(38,0){\makebox(0,0){
\begin{picture}( 20, 40)(- 10,- 20)
\thicklines
\put(0,0){\oval(10,30)[ r]}
\thinlines
\put(0,10){\oval(10,10)[ tl]}
\put(0,-10){\oval(10,10)[ bl]}
\put(- 5,- 6){\line(0,1){4}}
\put(- 5,2){\line(0,1){4}}
\end{picture}
}}
\put(- 20,0){\makebox(0,0){ ${\cal M} $}}
\put( 20,0){\makebox(0,0){ ${\cal M}' $}}
\put(0,0){\makebox(0,0){\scriptsize $E$}}
\put(0,- 40){\makebox(0,0){ (a)}}
\end{picture}
\centering
\begin{picture}(200,100)(- 100,- 50)
\thicklines
\put(0,0){\oval(60,60)}
\put(0,- 30){\line(0,1){ 60}}
\put(- 15,- 8){\makebox(0,0){ ${\cal M} $}}
\put( 15,- 8){\makebox(0,0){ ${\cal M}' $}}
\put(5,20){\makebox(0,0){\scriptsize $E$}}
\put(0,- 40){\makebox(0,0){(b)}}
\end{picture}

\centering
\begin{picture}(200,100)(- 100,- 50)
\thicklines
\put(0,0){\oval(70,50)}
\put(0,3){\oval(34,20)[b]}
\put(0,0){\oval(30,14)[t]}
\put(0,16){\oval(8,18)[r]}
\thinlines
\put(0,16){\makebox(0,0){
\begin{picture}(10,20)(- 5,- 10)
\put(0,5){\oval(8,8)[tl]}
\put(0,-5){\oval(8,8)[bl]}
\put(- 4,- 2){\line(0,1){4}}
%\put(4,-2){\vector(0,1){4}}
\end{picture}
}}
\put(0,- 16){\makebox(0,0){ ${\cal M} $}}
\put(0,16){\makebox(0,0){\scriptsize $E$}}
\put(0,- 40){\makebox(0,0){ (c)}}
\end{picture}
\begin{picture}(200,100)(- 100,- 50)
\thicklines
\put(0,0){\oval(50,50)}
\put(0,5){\oval(22,12)[ t]}
\put(0,-5){\oval(22,12)[ b]}
\put(11,- 5){\line(0,1){ 10}}
\put(-11,- 5){\line(0,1){ 10}}
\put(0,11){\line(0,1){14}}
\put(5,18){\makebox(0,0){\scriptsize $E$}}
\put(0,-18){\makebox(0,0){ ${\cal M} $}}
\put(0,- 40){\makebox(0,0){ (d)}}
\end{picture}
\caption[x]{\footnotesize Gluing manifolds along a circle and a
segment}
\label{f:gluing}
\end{figure}
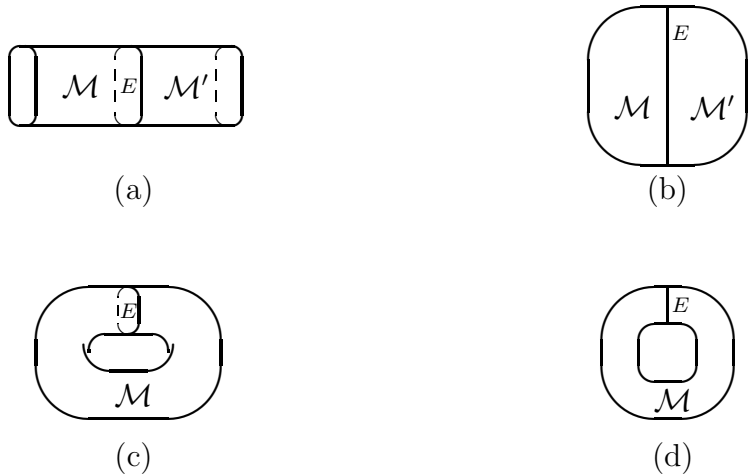

The next set of formulae we will need describe how
partition functions on two manifolds with boundaries can be glued
together to form a partition function on the manifold formed by
joining the two manifolds along a common edge.  In general, if two
manifolds ${\cal M} $ and ${\cal M}' $ are joined along a common edge
$E$, the partition function of the combined manifold
\begin{equation}
{\cal N}  = {\cal M}  \bigcup_E {\cal M}'
%\label{eq:}
\end{equation}
 is given by taking the product of the partition functions of ${\cal
M} $ and ${\cal M}' $ and integrating over the holonomy of the gauge
field along the edge $E$.  There are two cases of interest; the first
is when the edge $E$ is a closed 1-manifold homeomorphic to $S^1$
(Figure
\ref{f:gluing}a), the
second case is when $E$ is a line segment with boundary (Figure
\ref{f:gluing}b).

It is also necessary to consider the case where a manifold ${\cal M}$
is glued to itself along an edge that appears twice (with opposite
orientation) in the boundary of ${\cal M}$ (see
Figure~\ref{f:gluing}c,d).
In the basis of irreducible characters, the gluing formulae are
straightforward.  In the case of gluing along a circle, the relevant
result is the orthonormality of the characters
\begin{equation}
\int d U \chi_R (U) \chi_{R'} (U^{\dagger}) = \delta_{R,R'}.
\label{eq:circle}
\end{equation}
In the case of gluing along a segment, the formula is
\begin{equation}
\int d U \chi_R (VU) \chi_{R'} (U^{\dagger}W) =
\frac{\delta_{R,R'}}{\dim R}\chi_{R} (VW).
\label{eq:edge}
\end{equation}

We wish to convert (\ref{eq:circle}) and (\ref{eq:edge}) into
gluing formulae for partition functions expressed in terms of the
$\Upsilon_\sigma$ functions.
By using the Frobenius relations, the gluing formula on a circle
follows immediately; we have
\begin{eqnarray}
\int d U \Upsilon_\sigma (U) \Upsilon_\tau (U^{\dagger}) & = &
\int d U \sum_{R,R'} \chi_{R} (\sigma) \chi_{R'} (\tau)
\chi_{R} (U) \chi_{R'} (U^{\dagger})\nonumber\\
&  = &   \sum_{R}\chi_{R} (\sigma) \chi_{R} (\tau)
 =  \frac{\delta_{T_\sigma,T_\tau} n!}{| T_\sigma |}
\label{eq:circleupsilon},
\end{eqnarray}
where $T_\sigma$ is the conjugacy class of elements in the symmetric
group that contains $\sigma$.
This formula can be interpreted in the following fashion: Assume we
have two manifolds ${\cal M}$ and ${\cal M'}$, each of which has a
boundary component $E \approx S^1$.  Because we are working in a
single sector
of
the theory, the orientation of $E$ is taken to be opposite on the two
manifolds.  If the partition functions on ${\cal M}$ and ${\cal M'}$
are given by sums over covers with  $I$ and $I'$ $\Omega$-points
respectively, then we can combine the partition functions in a term
by
term fashion through (\ref{eq:circleupsilon}).  Given two $n$-fold
covers of
${\cal M}$ and ${\cal M'}$ whose boundary components above $E$ are
described by the conjugacy classes $T_\sigma$ and $T_\tau$, the
covers
can only be joined along $E$ when $T_\sigma = T_\tau$.  When these
conjugacy classes are equal, the number of ways that the covering
spaces can be glued together is exactly the number $C_\sigma$ of
elements of
$S_n$ that commute with an element $\sigma \in T_\sigma$.  This
number is exactly $ C_\sigma =n!/|T_\sigma |$.  Every cover of the
joined
manifold ${\cal N} $ can be formed in exactly such a fashion by
gluing
together covers of ${\cal M}$ and ${\cal M'}$.
Thus, (\ref{eq:circleupsilon})
states that  the partition function for ${\cal  N}$ is given by
taking
the sum over covers of ${\cal N} $ that are formed by joining
together covers of ${\cal M}$ and ${\cal M'}$ in all possible ways
along the common boundary.  Note that in this gluing process, no new
$\Omega$-points are introduced.  The argument given here also shows
that when a manifold ${\cal M}$ is glued to itself along a circle
$E\approx
S^1$, as in Figure~\ref{f:gluing}c,  the partition function of the
new manifold
is given by gluing
all covers of ${\cal M}$ together along $E$, without introducing new
$\Omega$-points.

Generalizing the  formula for gluing along a circle to the coupled
theory is again a straightforward application of the generalized
Frobenius relations.  As shown in Appendix A, this gluing formula is
given by
\begin{equation}
\int d U \Upsilon_{\bar{\tau}\sigma} (U, U^{\dagger})
\Upsilon_{\bar{\tau}'\sigma'}
(U, U^{\dagger}) = \delta_{T_\sigma, T_{\tau'}} \delta_{T_{\sigma'}, T_{\tau}}
C_\sigma C_\tau.
\label{eq:circleupsilon2}
\end{equation}
Again, this formula has the interpretation that when two manifolds
are
glued together along a circle (or a manifold is glued to itself), the
partition function is joined in the string picture by gluing all
covers together in all possible ways, without introducing new
$\Omega$-points.

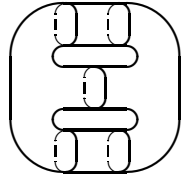
\begin{figure}
\centering
\begin{picture}(200,100)(- 100,- 50)
\thicklines
\put(0,12){\oval(32,8)}
\put(0,12){\oval(64,40)[t]}
\put(0,-12){\oval(32,8)}
\put(0,-12){\oval(64,40)[b]}
\put(- 32,- 12){\line(0,1){24}}
\put( 32,- 12){\line(0,1){24}}
\thinlines
\put(0,0){\makebox(0,0){
\begin{picture}(10,20)(- 5,- 10)
\put(0,4){\oval(8,8)[tl]}
\put(0,-4){\oval(8,8)[bl]}
\put(- 4,- 2){\line(0,1){4}}
\thicklines
\put(0,0){\oval(8,16)[r]}
\end{picture}
}}
\put(9,24){\makebox(0,0){
\begin{picture}(10,20)(- 5,- 10)
\put(0,4){\oval(8,8)[tl]}
\put(0,-4){\oval(8,8)[bl]}
\put(- 4,- 2){\line(0,1){4}}
\thicklines
\put(0,0){\oval(8,16)[r]}
\end{picture}
}}
\put(9, -24){\makebox(0,0){
\begin{picture}(10,20)(- 5,- 10)
\put(0,4){\oval(8,8)[tl]}
\put(0,-4){\oval(8,8)[bl]}
\put(- 4,- 2){\line(0,1){4}}
\thicklines
\put(0,0){\oval(8,16)[r]}
\end{picture}
}}
\put(- 11,24){\makebox(0,0){
\begin{picture}(10,20)(- 5,- 10)
\put(0,4){\oval(8,8)[tl]}
\put(0,-4){\oval(8,8)[bl]}
\put(- 4,- 2){\line(0,1){4}}
\thicklines
\put(0,0){\oval(8,16)[r]}
\end{picture}
}}
\put(- 11,- 24){\makebox(0,0){
\begin{picture}(10,20)(- 5,- 10)
\put(0,4){\oval(8,8)[tl]}
\put(0,-4){\oval(8,8)[bl]}
\put(- 4,- 2){\line(0,1){4}}
\thicklines
\put(0,0){\oval(8,16)[r]}
\end{picture}
}}
\end{picture}
\caption[x]{\footnotesize A genus 2 surface built from spheres with
boundary}
\label{f:glued}
\end{figure}

We observe in passing that the gluing formula
(\ref{eq:circleupsilon2}) can be used to construct the partition
function on a manifold of any genus by gluing together spheres with
1,2, and 3 boundary components.  These three objects, along with the
gluing formula (\ref{eq:circleupsilon2}), form a fundamental set of
vertices that can be viewed as yielding a set of ``Feynman rules'' for
the string theory.  The sphere with three boundary components (``pair
of pants'') is like a vertex; this partition function is given by a
sum over covering maps with one $\Omega^{-1}$-point.  The sphere with
two boundary components (the annulus or the cylinder) is like a
propagator.  This partition function is given by a sum over covering
maps with no twist points of either type.  The sphere with one
boundary component (the disk or cap), is analogous to a tadpole and
has a partition function given by a sum over covering maps with a
single $\Omega$-point.  Since gluing along a circle introduces no new
twist points, it is possible to glue together these fundamental
objects to form a genus $G$ Riemann surface, with a total of $2-2 G$
$\Omega$-points, where $\Omega^{-1}$-points are counted as $-1$.  Note
that the effect of an $\Omega^{-1}$-point is exactly to cancel a
single $\Omega$-point.  As an example, in Figure~\ref{f:glued} a genus
2 surface is built out of 2 vertices and 2 propagators for a total of
$- 2$ $\Omega$-points.  Note that the viewpoint expressed here differs
from the related familiar description of conformal and topological
theories, in that we are gluing together the target space and not the
string world sheet.

Now let us consider the formula needed to glue $\Upsilon$ functions
along a line segment.  The most general expression which will be of
use is a formula for the integral
\begin{equation}
\int dU \mixten{U}{i_1}{j_1}\cdots \mixten{U}{i_n}{j_n}
\mixten{U}{\dagger k_1}{l_1} \cdots\mixten{U}{\dagger k_n}{l_n}.
\end{equation}
{}From the symmetry of this expression under permutations of the
indices, it is clear that the integral is given by a sum over Wick
contractions with various permutations $\rho$ and unknown
coefficients
$W_\rho$,
\begin{equation}
\int dU \mixten{U}{i_1}{j_1}\cdots \mixten{U}{i_n}{j_n}
\mixten{U}{\dagger k_1}{l_1} \cdots\mixten{U}{\dagger k_n}{l_n}=
\sum_{\rho, \sigma}W_\rho
\mixten{\delta}{i_1}{l_{\sigma_1}}
\mixten{\delta}{k_{(\rho\sigma)_1}}{j_1}
\cdots
\mixten{\delta}{i_n}{l_{\sigma_n}}
\mixten{\delta}{k_{(\rho\sigma)_n}}{j_n},
\label{eq:unknown}
\end{equation}
where $\sigma_i$ is the image of $i$ under the permutation $\sigma$
of
the integers $1, \ldots, n$.  To determine the coefficients $W_\rho$,
we consider the related integral
\begin{eqnarray}
\int d U  ({\rm Tr}\;VU)^n  ({\rm Tr}\;U^{\dagger} W)^n  & = &
\int d U \sum_R d_R \chi_{R} (VU) \sum_{R'} d_{R'} \chi_{R'}
(U^{\dagger} W)
\nonumber\\
=  \sum_R {d_R^2 \over \dim R} \chi_{R} (VU) &=&
  \sum_{R}\frac{d_R}{N^n}
\sum_{\rho}\chi_{R}  \left( \Omega_n^{-1} \rho \right)
\Upsilon_{\rho} (VW)   \nonumber\\
 = \frac{n!}{N^n}  \sum_{\rho}\delta \left(
\Omega_n^{-1} \rho \right)
\Upsilon_{\rho} (VW)
 &=&   \frac{n!}{N^n}  \sum_{\rho}
w_\rho
\Upsilon_{\rho}(VW),
\label{eq:edgeintegral}
\end{eqnarray}
where we have used (\ref{eq:dimthree}).
Comparing (\ref{eq:edgeintegral}) to (\ref{eq:unknown}), we see that
\footnote{A similar formula was derived  by S. Samuel \cite{samuel} .}
\begin{equation}
W_\rho = \frac{w_\rho}{N^n} ,
%\label{eq:}
\end{equation}
%\sum_{R}\frac{d_R}{N^n}  \left(
%\sum_{\sigma \in S_n}
%\frac{\chi_{R} (\sigma)}{d_R N^{n - K_\sigma}}  \right)^{-1}
%\sum_{\rho}\chi_{R} (\rho)
%\Upsilon_{\rho} (VW) \nonumber\\
\begin{equation}
\int dU \mixten{U}{i_1}{j_1}\cdots \mixten{U}{i_n}{j_n}
\mixten{U}{\dagger k_1}{l_1} \cdots\mixten{U}{\dagger k_n}{l_n}=
\sum_{\rho, \sigma}\frac{w_\rho}{N^n}
\mixten{\delta}{i_1}{l_{\sigma_1}}
\mixten{\delta}{k_{(\rho\sigma)_1}}{j_1}
\cdots
\mixten{\delta}{i_n}{l_{\sigma_n}}
\mixten{\delta}{k_{(\rho\sigma)_n}}{j_n}.
\label{eq:segment}
\end{equation}
This equation can be interpreted in terms of gluing together covering
maps in a simple fashion.  If the only term in (\ref{eq:segment})
were
the leading term corresponding to $\rho$ being the identity
permutation, then the integral would simply give the well-known Wick
contraction of all edges, that corresponds to gluing the edges of
the
covering spaces together in all possible ways \cite{previous}.  The
existence of the extra factor of $\rho$, however, indicates
that the edges on one side or the other must be allowed to twist in
an
arbitrary fashion.  Geometrically, this twisting is exactly described
by putting an extra $\Omega^{-1}$-point on one side of the edge
before
gluing, and then applying the simple Wick contraction prescription.
Thus, we can describe the gluing together of the manifolds ${\cal M}$
and ${\cal M'}$ along a common segment $E$ by inserting an extra
$\Omega^{-1}$-point in ${\cal M}$, and then gluing together all
covers
of ${\cal M}$ (including the extra twist) with all covers of ${\cal
M'}$.  Similarly, when a manifold ${\cal M}$ is glued to itself along
a segment $E$, the partition function of the new manifold is formed
by
inserting an extra $\Omega^{-1}$-point in ${\cal M}$ before gluing
the
covering spaces together by a Wick contraction.

We have thus completed a description of all gluing processes in a
single sector of the theory.  When we consider the full theory, it is
again possible to calculate the result of gluing together manifolds
along an edge segment using (\ref{eq:segment}); however, this process
does not have such a simple description as in a single sector.  The
leading order terms, of course, combine according to the Wick
contraction; however, the lower order terms combine in a  more
complicated fashion since the basis functions
$\Upsilon_{\overline{\tau}\sigma} (U, U^{\dagger})$ contain both
factors of $U$
and
$U^{\dagger}$.  In order to achieve a simple geometric
understanding of this formula in the coupled theory, one must either
contract the indices of (\ref{eq:segment}) with other edges around a
boundary, or one must introduce a complicated set of orthogonal
polynomials in $U$ and $U^{\dagger}$ with an arbitrary number of
indices.

It is appropriate at this point to comment on past attempts to derive a string
representation of QCD. This effort has a long history \cite{history}. Most of
the attempts were based on the strong coupling expansion of the Wilson lattice
gauge theory, which to leading order can be written as a sum over surfaces with
simple weights. One expanded the Wilson action, $ \exp[-1/g^2  {\rm  Tr}
\left(U_P+U^{\dagger}_P \right) ] $,  where $ U_P $ is the holonomy around an
elementary plaquette of the lattice,  in powers of $1/g^2$. Then one used the
integral (\ref{eq:segment}) to join the plaquettes to form a surface. The
culmination of this enterprise was the work of Kazakov \cite{kaz},  of
Kostov  \cite{kos}  \cite{kos2} and  of O'Brien and Zuber \cite{obrien}.
They gave rules that expressed the strong coupling expansion as a complicated
sum over surfaces with an infinite number of local factors. Since this
approach was based on an expansion about infinite coupling the nature of the
continuum limit  was very unclear.  In addition, the discovery of large $N$
phase transitions \cite{growit} raised additional suspicions about the
relevance of these results to the continuum theory.

	The approach that we are following is very different. As proposed in
\cite{gross}, we are  using the exact solution of two dimensional QCD to
construct an equivalent string theory, which we then hope to continue to higher
dimensions. The exact lattice representation  we employ is one for which the
coupling is always in the weak coupling phase. The strong coupling limit of the
Wilson action is not accessible using this formulation of the theory. In higher
dimensions of course this method is not available. If we understood enough
about QCD to derive its string theory representation then the string theory
would be superfluous. The reason that we need such a representation is
precisely because we do not have analytic control over the theory in the
infra-red. Instead we hope to employ the lessons of two dimensions and  wisdom
of string theory to guess the QCD string theory in four dimensions.

\section{Wilson loops}
\setcounter{equation}{0}
\baselineskip 18.5pt

We now wish to calculate the vacuum expectation values (VEV's) of
Wilson loops in QCD$_2$ and to relate these VEV's to string theory
maps.  We derive in this section a general formula that can be used
to compute vacuum expectation values of arbitrary products of Wilson
loops on an arbitrary 2-manifold in a single sector of the theory.
We
will prove this formula by using the Wilson loops to cut the manifold
into pieces, each of which is a manifold with boundary, and then
gluing the pieces together along the loops using the gluing formulae
{}from the previous section.  We find that the vacuum expectation value
of any product of Wilson loops is given by a sum similar to
(\ref{eq:general}), where each of the regions bounded by the loops
contains some number of $\Omega$-points or $\Omega^{-1}$-points, and
where the covering space has a boundary that covers the Wilson loops
in a fashion that is determined by the labels on the loops.  We do
not give a general formula here for the expectation values of Wilson
loops in the coupled theory.  We do, however, describe an algorithm
by
means of which an arbitrary such expectation value can be computed.
Several simple examples of Wilson loop calculations in the coupled
theory are described in section 7.  To derive a general formula for
the
full theory would involve a similar analysis to that performed in
this section; however, the technical details become far more complex.
Thus, for most of this section, we will confine our discussion to a
single sector of the theory.

As mentioned in the Introduction, in the string theory the natural
type of label to associate with a Wilson loop $\gamma$ is an integer
$k$ and a permutation $\sigma \in S_k$.  The VEV of the Wilson loop
with this label is defined to be $\langle \Upsilon_\sigma (U)
\rangle$, where $U$ is the holonomy of the gauge field around the
Wilson loop.  It should be clear from the discussion of manifolds
with
boundaries in the previous section that these are the most natural
observables from the point of view of string maps, since the
functions
$\Upsilon_\sigma$ are the natural basis for the Hilbert space in the
string theory picture.  These observables are related to the more
familiar vacuum expectation values of $\langle \chi_{R} (U)\rangle$
by
the Frobenius relations.  In a single sector of the theory, it is
sufficient to consider VEV's of functions $\Upsilon_\sigma (U)$; in
the coupled theory, one would like to compute VEV's of arbitrary
functions $\Upsilon_{\overline{\tau}\sigma} (U, U^{\dagger})$.  We
will denote VEV's in the orientation-preserving sector by
$\langle\Upsilon_\sigma (U)\rangle_+$.

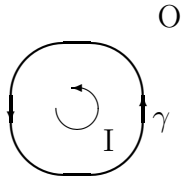
\begin{figure}
\centering
\begin{picture}(200,100)( - 100,- 50)
\thicklines
\put(0,0){\oval(50,50)}
\thinlines
\put(0,0){\oval(16,16)[b]}
\put(0,0){\oval(16,16)[tr]}
\put(1,8){\vector(-1,0){3}}
\put(25,- 5){\vector(0,1){10}}
\put(-25, 5){\vector(0,-1){10}}
\put(12,-12){\makebox(0,0){I}}
\put(35,35){\makebox(0,0){O}}
\put(30,-5){\makebox(0,0){ $\gamma$}}
\end{picture}
\caption[x]{\footnotesize the inside (I) and outside (O) of an
oriented closed curve $\gamma$}
\label{f:orientation}
\end{figure}

We assume that the manifold ${\cal M}$ is provided with a preferred
orientation.  If we consider a local region of the manifold to be
embedded in Euclidian 3-space, we can relate the orientation of
${\cal
M}$ to a normal vector field $\hat{n} (z)$.  Each Wilson loop is also
associated with some orientation, according to the direction in which
the gauge field is integrated.  Thus, it is possible to distinguish
the two sides of a Wilson loop at a point.  We will use the labels
``inside'' and ``outside'' to denote these two sides.  If
$\hat{\gamma}(z)$ is the oriented tangent vector to $\gamma$ at a
point $z$, then the vector $\hat{n} (z)\times \hat{\gamma}(z)$ points
into the region on the inside of $\gamma$ (see
Figure~\ref{f:orientation}).  When we are working in a single sector
of the gauge theory on a manifold with boundary, we are restricting
to
covering spaces that lie on a specific side of the boundary curves
$E_i$.  We will set the convention that in the orientation-preserving
sector of the theory, the covering spaces must lie on the ``outside''
of the boundary curves.  Although this choice of notation may seem
odd, it leads to a more intuitive description of Wilson loops; the
orientation-preserving sheets of a cover that are bounded by a Wilson
loop always lie on the inside of the loop.

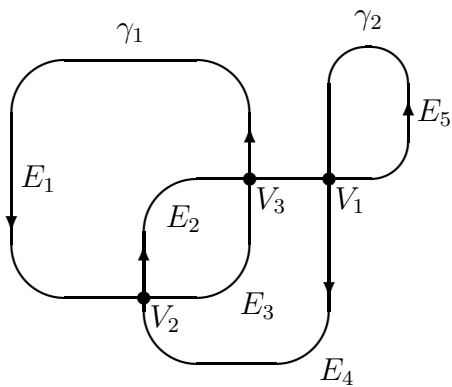
\begin{figure}
\centering
\begin{picture}(200,150)( - 100,- 55)
\thicklines
\put(- 30,20){\oval(90,90)}
\put(15,20){\vector(0,1){20}}
\put(- 75,20){\vector(0,-1){20}}
\put(10,- 15){\oval(70,70)[b]}
\put(-25,- 15){\vector(0,1){10}}
\put(45,- 15){\vector(0,-1){10}}
\put(10,- 15){\oval(70,70)[tl]}
\put(45, -15){\line(0,1){60}}
\put(10,20){\line(1,0){50}}
\put(60,45){\oval(30,50)[t]}
\put(75,45){\vector(0,1){5}}
\put(60,45){\oval(30,50)[rb]}
\thinlines
\put(- 65,20){\makebox(0,0){$E_1$}}
\put(- 10,5){\makebox(0,0){$E_2$}}
\put(18,- 28){\makebox(0,0){$E_3$}}
\put(48,- 53){\makebox(0,0){$E_4$}}
\put(85,45){\makebox(0,0){$E_5$}}
\put(- 30,75){\makebox(0,0){$\gamma_1$}}
\put(60,80){\makebox(0,0){$\gamma_2$}}
\put(45,20){\circle*{5}}
\put(53,12){\makebox(0,0){$V_1$}}
\put(- 25,- 25){\circle*{5}}
\put(- 17,- 33){\makebox(0,0){$V_2$}}
\put(15,20){\circle*{5}}
\put(23,12){\makebox(0,0){$V_3$}}
\end{picture}
\caption[x]{\footnotesize A graph $\Gamma$ formed by Wilson
loops $\gamma_1, \gamma_2$ with edges $E_a$, vertices $V_b$}
\label{f:graph}
\end{figure}

Given an arbitrary set of Wilson loops $\gamma_i$ carrying the labels
$(k_i, \sigma_i \in S_{k_i})$, we are interested in computing the VEV
\begin{equation}
\langle W_{\{\gamma_i, \sigma_i\}} \rangle_+
= \langle \prod_{j} \Upsilon_{\sigma_j} (U_j) \rangle_+,
%\label{eq:}
\end{equation}
where $U_j$ is the holonomy around the $j$th Wilson loop $\gamma_j$.
We will assume that the Wilson loops do not have intersection points
where more than two loops intersect (including self-intersections).
Wilson loops that do not satisfy this criterion can be computed by
taking the limit of a set of Wilson loops that do satisfy the
criterion.  With this assumption, we can consider the Wilson loops
$\gamma_j$ to form a graph $\Gamma$ on ${\cal M}$, with vertices
$V_1,
\ldots, V_v$ at intersection points and edges $E_1, \ldots, E_e$
connecting the vertices along the Wilson loops (see
Figure~\ref{f:graph}).  By cutting along the Wilson loops, ${\cal
M}\setminus (\bigcup \gamma_i)$ is divided into a disjoint union of
connected regions $M_j$.  If we denote the areas of these regions by
$A_j$, then the VEV of $W_{\{\gamma_i, \sigma_i\}}$ is a function of
the $A_j$.

We note in passing that the VEV of a Wilson loop with an arbitrary
label $(k, \sigma)$ can be computed as the limit of the VEV of a
Wilson loop in the fundamental representation ($k =1$) as the areas
of
certain regions go to 0.  For example, the VEV of the trivial Wilson
loop on the sphere (Figure~\ref{f:loops1}(a)) with the label $(2,
\sigma_2 = [2^1])$ is equal to the VEV of the double Wilson loop on
the sphere as the intermediate area goes to 0
(Figure~\ref{f:loops1}(b) as $A_3 \rightarrow 0$).  Nonetheless, we
will now give the formula for the expectation value of a product of
loops with arbitrary labels.

\begin{figure}
\centering
\begin{picture}(200,150)( - 100,- 55)
\thicklines
\put(- 30,20){\oval(90,90)}
\put(15,20){\vector(0,1){20}}
\put(- 75,20){\vector(0,-1){20}}
\put(10,- 15){\oval(70,70)[b]}
\put(-25,- 15){\vector(0,1){10}}
\put(45,- 15){\vector(0,-1){10}}
\put(10,- 15){\oval(70,70)[tl]}
\put(45, -15){\line(0,1){60}}
\put(10,20){\line(1,0){50}}
\put(60,45){\oval(30,50)[t]}
\put(75,45){\vector(0,1){5}}
\put(60,45){\oval(30,50)[rb]}
\thinlines
\put(- 45,35){\makebox(0,0){\footnotesize $M_1 (1)$}}
\put(- 5,0){\makebox(0,0){\footnotesize  $M_2 (0)$}}
\put(10,- 35){\makebox(0,0){\footnotesize $M_3 (1)$}}
\put(70,- 20){\makebox(0,0){\footnotesize $M_4 (-1)$}}
\put(60,45){\makebox(0,0){\footnotesize $M_5 (1)$}}
\put(- 30,75){\makebox(0,0){$\gamma_1$}}
\put(60,80){\makebox(0,0){$\gamma_2$}}
\put(45,20){\circle*{5}}
\put(- 25,- 25){\circle*{5}}
\put(15,20){\circle*{5}}
\end{picture}
\caption[x]{\footnotesize Numbers of $\Omega$-points in each region
$M_j (I_j)$}
\label{f:graph2}
\end{figure}
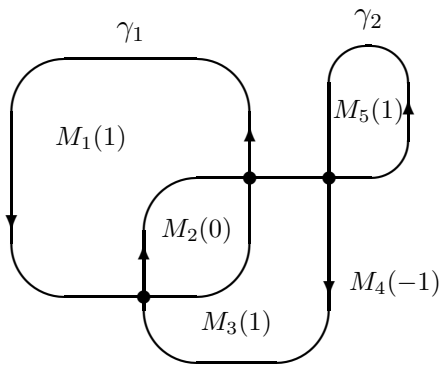

The result is computed as follows: to each region $M_j$ we assign a
number of $\Omega$-points, $I_j$, according to the parts of the graph
$\Gamma$ that lie on the boundary of $M_j$.  Specifically, each
region $M_j$ contributes $2 -2G_j - b_j$ to $I_j$, where $G_j$ is the
genus of $M_j$ and $b_j$ is the number of boundary components.  In
addition, each edge of $\Gamma$ contributes an $\Omega^{-1}$-point to
the region on its inside, and each vertex of $\Gamma$ contributes an
$\Omega$-point to the region that lies on the inside of both edges
that pass through that vertex.  As an example, in
Figure~\ref{f:graph2}, the numbers of $\Omega$-points are shown for
each region of the sphere after cutting along the Wilson loops from
Figure~\ref{f:graph}.

We define $\Sigma_+ ({\cal M};\{\gamma_i, \sigma_i\})$ to be the set
of orientation-preserving covers of ${\cal M}$ with the usual
singularities, $I_j$ $\Omega$-points in the region $M_j$, and a
boundary that covers the Wilson loops $\gamma_i$ with the
permutations $\sigma_i$.  The expectation value of the product of
Wilson loops can be expressed as a sum over the covers in this set,
\begin{equation}
\langle W_{\{\gamma_i, \sigma_i\}} \rangle_+ =
\int_{\Sigma_+ ({\cal M};\{\gamma_i, \sigma_i\} )} d \nu
\;  e^{- \frac{n \lambda A}{2} }
\frac{(-1)^{i} N^{2 -2 g- \beta}}{ |S_\nu |}  \prod_{j}
\prod_{k=1 }^{\max (0, - I_j)} (-1)^{x_{(j,k)}},
\label{eq:wilson}
\end{equation}
where everything is defined as in (\ref{eq:b}), and $x_{(j,k)}$ is
the
number of twists at the $k$th $\Omega^{-1}$-point in the region
$M_j$.

\begin{figure}
\centering
\begin{picture}(200,100)(- 100,- 50)
\put(- 90,0){\makebox(0,0){
\centering
\begin{picture}(200,100)(- 100,- 50)
\thicklines
\put(- 30,0){\vector(1,0){60}}
\put(0,- 30){\vector(0,1){60}}
\put(7,- 10){\makebox(0,0){ $V$}}
\put(0,0){\circle*{5}}
\put(- 20,20){\makebox(0,0){$M_1$}}
\put(20,20){\makebox(0,0){$M_4$}}
\put(- 20,-20){\makebox(0,0){$M_2$}}
\put(20,-20){\makebox(0,0){$M_3$}}
\put(6,- 28){\makebox(0,0){$\gamma_1$}}
\put(- 28,5){\makebox(0,0){ $\gamma_2$}}
\end{picture}}}
\thinlines
\put(- 25,0){\vector(1,0){ 30}}
\put(90,0){\makebox(0,0){
\begin{picture}(200,100)(- 100,- 50)
%\put(0,5){\oval(40,30)[t]}
%\put(0,-5){\oval(40,30)[b]}
%\put(20,- 5){\line(0,1){10}}
%\put(-20,- 5){\line(0,1){10}}
\put(0,0){\circle{32}}
\put(5,5){\oval(40,40)[tr]}
\put(-5,5){\oval(40,40)[tl]}
\put(5,-5){\oval(40,40)[br]}
\put(-5,-5){\oval(40,40)[bl]}
\put(5,25){\line(0,1){30}}
\put(25,5){\line(1,0){30}}
\put(-5,25){\line(0,1){30}}
\put(-25,5){\line(-1,0){30}}
\put(5,-25){\line(0,-1){30}}
\put(25,-5){\line(1,0){30}}
\put(-5,-25){\line(0,-1){30}}
\put(-25,-5){\line(-1,0){30}}
\put(-40,40){\makebox(0,0){$M_1$}}
\put(40,40){\makebox(0,0){$M_4$}}
\put(-40,-40){\makebox(0,0){$M_2$}}
\put(40,-40){\makebox(0,0){$M_3$}}
\put(0,0){\makebox(0,0){$D$}}
\thicklines
\put(- 50,0){\line(1,0){27}}
\put(0,0){\oval(46,46)[ b]}
\put(23,0){\vector(1,0){27}}
\put(0,- 50){\line(0,1){31}}
\put(0,0){\oval(38,38)[ r]}
\put(0,19){\vector(0,1){31}}
\put(- 57,0){\makebox(0,0){$\gamma_2$}}
\put(1,- 57){\makebox(0,0){$\gamma_1$}}
\end{picture}
}}
\end{picture}

\caption[x]{\footnotesize a disk $D$ at the
intersection $V$ of two Wilson loops}
\label{f:disks}
\end{figure}
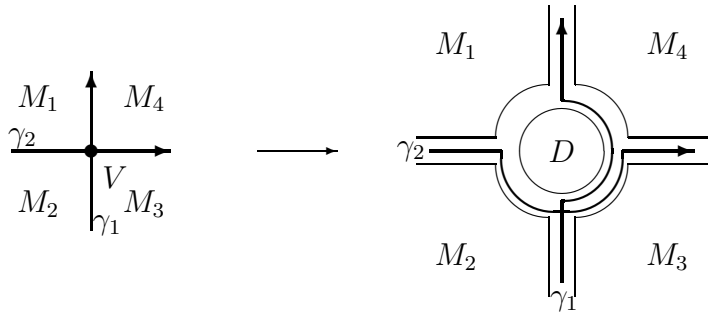

We will now prove this assertion, by cutting the manifold along the
Wilson loops and gluing together using the formulae from the previous
section.  In order to make the calculation clearer, it will be useful
to define a set of infinitesimal disks $D_1, \ldots, D_v$ around the
vertices $V_i$.  We will assume that the Wilson loops are deformed so
that the disk $D_i$ lies on the inside of the Wilson loops passing
through the vertex $V_i$ (see figure~\ref{f:disks}).
The gluing process can then be performed in the following order:
first
the partition functions over the regions $M_j$ are glued along the
edges $E_1, \ldots, E_e$ using (\ref{eq:segment}), then the disks
$D_i$ are glued into the resulting manifold using
(\ref{eq:circleupsilon}).

When we begin this gluing process, the partition function over each
region $M_j$ is described by a sum over covers of $M_j$ with $2 -2G_j
- b_j$ $\Omega$-points, as was demonstrated in the previous section.
Each time we glue together two regions along an edge $E$, we must
apply the gluing formula (\ref{eq:segment}).  Since some Wilson loop
$\gamma_i$ passes along $E$, we must include an extra $k_i$ copies of
$U$ in the left-hand side of this gluing formula (we assume that the
edge is oriented consistently with $\gamma_i$).  Clearly, this
integral will only be non-zero when the number of sheets of the cover
on the inside of $E$ (corresponding to factors of $U^{\dagger}$)
exceeds the number of sheets on the outside (corresponding to factors
of $U$) by $k_i$.  Note that if the same region lies on both sides of
the edge $E$, this condition is never satisfied, and the Wilson loop
VEV must be 0.  We described the result of (\ref{eq:segment})
geometrically by observing that this gluing formula was equivalent to
twisting the boundaries on one side of the edge by an
$\Omega^{-1}$-point, then gluing in all possible ways according to
the
Wick contraction.  We can use this same geometric interpretation
here;
we must only ensure that the $\Omega^{-1}$-point does not mix edges
{}from the Wilson loop with edges on the boundary of the outside
region.
This is easily done by using the $\Omega^{-1}$-point to twist the
edges on the inside region (the factors of $U^{\dagger}$).  This is
equivalent to placing an extra $\Omega^{-1}$-point in the inside
region, and accounts for the edge contribution to $I_j$ given above.

After we have glued the regions along all edges $E_i$ of $\Gamma$, it
remains to glue in the disks $D_i$.  Each disk $D$ is associated with
a single $\Omega$-point.  Because we have deformed the Wilson loops
such that $D$ is on the inside of all loops, the number of sheets in
the cover of the boundary of $D$ (the number of factors of $U$) is
constant around the boundary.  Thus, we may apply
(\ref{eq:circleupsilon}), and we find that the disk $D$ is added with
no extra twists; the single $\Omega$-point from $D$ can be associated
with the region on the inside of both Wilson loops (in
Figure~\ref{f:disks}, $M_1$).

Thus, we have proven (\ref{eq:wilson}).  This gives the general
formula for the vacuum expectation value of an arbitrary product of
Wilson loops in a single sector of the theory.  The natural next
problem is to generalize this result to the full, coupled theory.
Although the vacuum expectation value of a specific Wilson loop in
the
full theory can be calculated using the formulae (\ref{eq:segment})
and (\ref{eq:circleupsilon2}), we do not yet have such a simple
description of the geometry of this computation.  The difficulties
that arise in the coupled theory arise from two related phenomena.
First, a Wilson loop in the coupled theory can either bound an
orientation-preserving sheet on the inside, or an
orientation-reversing sheet on the outside.  Second, the extra terms
in the functions $\Upsilon_{\overline{\tau}\sigma} (U, U^{\dagger})$,
that arise from the necessity to suppress folds that would arise
{}from contracting factors of $U$ and $U^{\dagger}$ in the function
$\Upsilon_\sigma (U)\Upsilon_\tau (U^{\dagger})$, make computations
significantly more difficult.

The first place in the analysis where the coupled theory is more
complicated is when one glues regions together along the edges $E_i$
of $\Gamma$.  If a Wilson loop bounds sheets on both sides, it is
impossible to avoid having the $\Omega^{-1}$-point mix the Wilson
loop
with edges of the boundaries.  This problem can be avoided by
restricting attention to Wilson loops carrying the fundamental
representation of $SU(N)$ $(k =1)$.  As noted above, the expectation
values of all Wilson loops with arbitrary labels can be related to
such fundamental loops, so this is not a significant restriction.
With this constraint, each edge of the Wilson loop only bounds any
given cover on a single side, so the $\Omega^{-1}$-points arising
{}from
edges can be distributed as in the single sector theory.  Note,
however, that now the quantities $I_j$ depend upon the distribution
of
sheets in the cover.

\begin{figure}
\centering
\begin{picture}(200,100)(- 100,- 50)
\put(0,0){\circle{32}}
\thicklines
\put(8,8){\oval(34,34)[tr]}
\put(-8,8){\oval(34,34)[tl]}
\put(8,-8){\oval(34,34)[br]}
\put(-8,-8){\oval(34,34)[bl]}
\put(8,25){\vector(-1,0){3}}
\put(25,8){\line(0,-1){3}}
\put(-25,8){\vector(0,-1){3}}
\put(8,-25){\line(-1,0){3}}
\put(-8, -25){\vector(1,0){3}}
\put( -25,-8){\line(0,1){3}}
\put(25,-8){\vector(0,1){3}}
\put(-8,25){\line(1,0){3}}
\thinlines
\put(5,25){\line(0,1){30}}
\put(25,5){\line(1,0){30}}
\put(-5,25){\line(0,1){30}}
\put(-25,5){\line(-1,0){30}}
\put(5,-25){\line(0,-1){30}}
\put(25,-5){\line(1,0){30}}
\put(-5,-25){\line(0,-1){30}}
\put(-25,-5){\line(-1,0){30}}
\put(-30,20){\makebox(0,0){$V$}}
\put(-40,40){\makebox(0,0){$M_1$}}
\put(30,20){\makebox(0,0){$Y$}}
\put(40,40){\makebox(0,0){$M_4$}}
\put(-30,-20){\makebox(0,0){$W$}}
\put(-40,-40){\makebox(0,0){$M_2$}}
\put(30,-20){\makebox(0,0){$X$}}
\put(40,-40){\makebox(0,0){$M_3$}}

\put(5,5){\makebox(0,0){\footnotesize $D$}}
\put(- 25,0){\vector(1,0){50}}
\put(0,-25){\vector(0,1){50}}
\put(- 32,0){\makebox(0,0){$i$}}
\put( 32,0){\makebox(0,0){$l$}}
\put(1,- 32){\makebox(0,0){$k$}}
\put(1, 32){\makebox(0,0){$j$}}
\end{picture}

\caption[x]{\footnotesize  Intersection point in the coupled theory}
\label{f:disks2}
\end{figure}
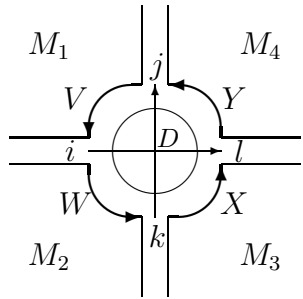

Even though by this means it is possible to understand the gluing
along edges simply, the final gluing of the disks raises more serious
technical difficulties.  Consider the disk $D$ depicted in
Figure~\ref{f:disks2}.  We have broken up the boundary of $D$ into
four segments with holonomies $V,W,X,Y$, corresponding to the
segments
that are contained in the boundaries of regions $M_1,M_2,M_3,M_4$.
After gluing together these regions, the partition function will be a
sum of terms whose dependence on $V,W,X,Y$ is described by a general
polynomial $\mixten{\psi}{ik}{jl}(V,W,X,Y)$, whose transformation
properties are indicated by the tensor indices.  The remaining
portions of the Wilson loops passing through this intersection are
given by $\mixten{\delta}{l}{i}$ and $\mixten{\delta}{j}{k}$ in the
limit where the area of the disk $D$ goes to 0.  To calculate the
effect of gluing in the disk, these Wilson loops must be contracted
with the indices of $\psi$, and the holonomies $V,W,X,Y$ must be
taken
to 1.

As an example of the type of situation that can arise in the coupled
theory, consider the case where a cover of ${\cal M}$ in the
neighborhood of $D$ has a single orientation-preserving cover in
region $M_1$ and a single orientation-reversing cover in region
$M_3$.
In such a case, the $\Omega$-point usually associated with $D$ cannot
be simply placed in one of the adjacent regions, since there is not a
unique adjacent region containing all the local sheets of the cover.
In this particular case, after the regions $M_1,M_2,M_3$, and $M_4$
have been glued together, the partition function will be proportional
to $\mixten{V}{i}{j}\mixten{X^{\dagger}}{k}{l}$.  This result follows
immediately from (\ref{eq:segment}).  To compute the contribution of
the disk, we compute
\begin{equation}
\lim_{V,X \rightarrow 1} \mixten{\delta}{l}{i}\mixten{\delta}{j}{k}
\mixten{V}{i}{j}\mixten{X^{\dagger}}{k}{l}= N.
\label{eq:simpletwist}
\end{equation}
This result can be interpreted as indicating that the sheets of
opposite orientation are connected at the vertex with a simple twist,
with no additional tubes or extra structure allowed.  We will compute
several examples of Wilson loops with this type of vertex in Section
7.

We conclude this section with a brief description of how the
contribution from a general vertex in the coupled theory can be
computed; we do not have a simple geometric description of this
computation, such as we have described in the single sector theory.
We will use the notation $U = YXWV$ for the total holonomy around the
disk $D$, and the abbreviation $\Delta$ for the product of Wilson
loop
tensors $\mixten{\delta}{l}{i}\mixten{\delta}{j}{k}$.  We will drop
the indices in most equations for clarity.  The most general form for
a function $\psi$ of the holonomies $V,W,X,Y$ around $D$ is given by
the functions
\begin{eqnarray}
A_{\sigma, \tau}^{s,t}(V,W,X,Y) & = &  \Upsilon_\sigma (U)
\Upsilon_\tau (U^{\dagger}) \mixten{(V U^s)}{i}{j}
\mixten{(X^{\dagger} Y^{\dagger}U^{t}Y)}{k}{l} \nonumber \\
B_{\sigma, \tau}^{s,t}(V,W,X,Y) & = &  \Upsilon_\sigma (U)
\Upsilon_\tau (U^{\dagger})
\mixten{(VU^s Y)}{i}{l}
\mixten{(X^{\dagger}Y^{\dagger} U^t)}{k}{j},
\end{eqnarray}
where $s,t$ range from $- \infty$ to $\infty$, and $\sigma, \tau$ are
arbitrary permutations.  If we compute the inner product on these
functions by contracting indices and integrating, we find that
\begin{eqnarray}
\int A_{\sigma, \tau}^{s,t} A_{\rho, \upsilon}^{\dagger r,u} & = &
\int d U \Upsilon_{\sigma + \upsilon} (U) \Upsilon_{\tau + \rho}
(U^{\dagger}) {\rm Tr}\; (U^{s - r}) {\rm Tr}\; (U^{t - u}) \\
\int A_{\sigma, \tau}^{s,t} B_{\rho, \upsilon}^{\dagger r,u} & = &
\int d U \Upsilon_{\sigma + \upsilon} (U) \Upsilon_{\tau + \rho}
(U^{\dagger}) {\rm Tr}\; (U^{s - r+ t - u}) \\
\int B_{\sigma, \tau}^{s,t} B_{\rho, \upsilon}^{\dagger r,u} & = &
\int d U \Upsilon_{\sigma + \upsilon} (U) \Upsilon_{\tau + \rho}
(U^{\dagger}) {\rm Tr}\; (U^{s - r}) {\rm Tr}\; (U^{t - u}) .
\end{eqnarray}
These integrals can all be explicitly evaluated using
(\ref{eq:circleupsilon}).

There is a natural partial ordering defined on these functions.  We
define $ (A,B)_{\sigma, \tau}^{s,t}\succ (A,B)_{\sigma, \tau}^{s,t}$
when the number of factors of each term $V,W,X,Y,V^{\dagger},
W^{\dagger}, X^{\dagger}, Y^{\dagger}$ in the first function is
greater than the number of corresponding terms in the second
function.
In this situation, we say that the first function ``dominates'' the
second.  Thus, for example,
\begin{equation}
A_{( \cdot), (\cdot)}^{-1,1}  = \mixten{(W^{\dagger} X^{\dagger}
Y^{\dagger})}{i}{j}\mixten{(WVY)}{k}{l} \succ
    B_{(\cdot), (\cdot)}^{0,0}
= \mixten{(V Y)}{i}{l}
\mixten{(X^{\dagger} Y^{\dagger})}{k}{j},
%\label{eq:}
\end{equation}
where we denote by $(\cdot)$ the trivial permutation on 0 elements.
On the other hand, considering the functions $A_{( \cdot), (\cdot)}^{
0,0} = \mixten{V}{i}{j}\mixten{(X^{\dagger})}{k}{l} $ and
$B_{(\cdot),
(\cdot)}^{0,1} = \mixten{(V Y)}{i}{l} \mixten{(WV)}{k}{j}$, we find
that neither function dominates the other.

In order to compute the effect of gluing in the disk $D$ when the
covering space of ${\cal M}\setminus D$ has a fixed number of sheets
in each region $M_i$, and a fixed permutation on the boundary of the
cover around $D$, one must use the following general procedure.
Associated with the boundary of the covering space, there is a
particular function $\psi=A_{\sigma, \tau}^{s,t}$ or $\psi
=B_{\sigma,
\tau}^{s,t}$ defined by the permutation on the boundary of the cover.
In the same way that the functions $\Upsilon_{\overline{\tau}\sigma}
(U, U^{\dagger})$, that provide a natural basis for the Hilbert space
in the theory without Wilson loops, were defined by orthonormalizing
the functions $\Upsilon_\sigma (U)\Upsilon_\tau (U^{\dagger})$, we
must orthonormalize the set of functions $A$ and $B$, to form a set
of
functions
\begin{eqnarray}
\alpha_{\sigma, \tau}^{s,t} & = & A_{\sigma, \tau}^{s,t}
+ \cdots\\
\beta_{\sigma, \tau}^{s,t} & = & B_{\sigma, \tau}^{s,t}
+ \cdots,
\end{eqnarray}
where the ellipses indicate subdominant terms, that are orthogonal to
all subdominant terms; i.e., that satisfy
\begin{equation}
\int \alpha_{\sigma, \tau}^{s,t} \psi^{\dagger} = 0, \; \; \;
\alpha_{\sigma, \tau}^{s,t} \succ \psi
%\label{eq:}
\end{equation}
\begin{equation}
\int \beta_{\sigma, \tau}^{s,t} \psi^{\dagger} = 0, \; \; \;
\beta_{\sigma, \tau}^{s,t} \succ \psi.
%\label{eq:}
\end{equation}
The function associated with the cover of ${\cal M}\setminus D$ will
then be the function $\alpha$ or $\beta$ with a leading term
consistent with the boundary of the cover.  To compute the effect of
gluing in the disk, we can then apply the formulae
\begin{eqnarray}
\lim_{V,W,X,Y \rightarrow 1}
\Delta \alpha_{\sigma, \tau}^{s,t}
& = &  N^{1 + K_\sigma + K_\tau}, \\
\lim_{V,W,X,Y \rightarrow 1}
\Delta \beta_{\sigma, \tau}^{s,t}
& = &  N^{2 + K_\sigma + K_\tau}.
\end{eqnarray}

We illustrate this general formalism with several examples.  First,
consider the covering described above, where $M_1$ has a single
orientation-preserving covering sheet and $M_3$ has an
orientation-reversing sheet.  The function describing
this cover is
\begin{equation}
\alpha_{(\cdot), (\cdot)}^{0,0} = A_{(\cdot), (\cdot)}^{0,0} =
\mixten{V}{i}{j}\mixten{X^{\dagger}}{k}{l}.
\end{equation}
Computing the effect of gluing in the disk, we have
\begin{equation}
\lim_{V,W,X,Y \rightarrow 1}
\Delta \alpha_{(\cdot), (\cdot)}^{ 0,0}
 =   N.
\label{eq:vertex1}
\end{equation}
As mentioned above, this has the effect of gluing together the
sheets of the cover with a simple twist and no tubes.

For another example, consider the situation where an
orientation-preserving sheet covers regions $M_1,M_2,M_4$ and an
orientation-reversing sheet covers regions $M_2,M_3,M_4$.  The
function associated with the boundary permutation is
$\mixten{( W^{\dagger} X^{\dagger}
 Y^{\dagger})}{i}{j}\mixten{(WVY)}{k}{l}=A_{( \cdot),
(\cdot)}^{-1,1}$.
Thus, we must find the function $\alpha_{( \cdot), (\cdot)}^{-1,1}$.
This function can be computed by writing a sum of all subdominant
terms with arbitrary coefficients
\begin{equation}
\alpha_{( \cdot), (\cdot)}^{-1,1} = A_{( \cdot), (\cdot)}^{-1,1}
+ a A_{(\cdot), (\cdot)}^{0,0}
+ b B_{(\cdot), (\cdot)}^{0,0}
+ c B_{(\cdot), (\cdot)}^{-1,1}.
\end{equation}
Taking the inner product with the subdominant terms, we have
\begin{equation}
\int \alpha_{(\cdot), (\cdot)}^{-1,1} (A_{(\cdot),
(\cdot)}^{0,0})^{\dagger}
= 1 + a N^2 + b N + c N = 0,
\end{equation}
\begin{equation}
\int \alpha_{(\cdot), (\cdot)}^{-1,1} (B_{(\cdot),
(\cdot)}^{0,0})^{\dagger}
= N + a N + b N^2 + c = 0,
\end{equation}
\begin{equation}
\int \alpha_{(\cdot), (\cdot)}^{-1,1} (B_{(\cdot),
(\cdot)}^{-1,1})^{\dagger}
= N + a N + b + c N^2 = 0.
\end{equation}
Thus, $a = 1/N^2,b = c = -1/N$,  so
\begin{equation}
\alpha_{( \cdot), (\cdot)}^{-1,1} = A_{( \cdot), (\cdot)}^{-1,1}
+  \frac{1}{N^2}  A_{(\cdot), (\cdot)}^{0,0}
- \frac{1}{N}   B_{(\cdot), (\cdot)}^{0,0}
- \frac{1}{N}  B_{(\cdot), (\cdot)}^{-1,1}.
\end{equation}
Taking the limit and contracting with the remaining Wilson loops,
\begin{equation}
\Delta
\alpha_{(\cdot), (\cdot)}^{-1,1}= -N(1 - \frac{1}{N^2} ).
\label{eq:vertex2}
\end{equation}
Thus, the effect of such a vertex is to join the sheets with a twist,
allowing an orientation-reversing tube between the sheets, and to
introduce an extra factor of $-1$.  We will consider an example of a
Wilson loop with such a vertex in Section 7; we  show there that this
result is in agreement with an earlier calculation by Kazakov using
loop equations\cite{kazakov}.

\section{Examples of partition functions}
\setcounter{equation}{0}
\baselineskip 18.5pt

We will now give a variety of examples of calculations in the string
theory picture.  We will primarily perform explicit calculations for
the terms corresponding to covering maps with small winding number,
corresponding to representations $\bar{S}R$ where $S$ and $R$ are
described by Young tableaux with a small number of boxes.  In this
section and the next, we will order the terms according to the number
of sheets in the covering spaces, rather than according to powers of
$1/N$. For $G>1$, for each power of $1/N$ it is only necessary to
compute the terms corresponding to covering spaces with a finite
number of sheets, since $g \geq 1 +n(G-1)$.

We recall from \cite{previous} that for a manifold ${\cal M}$ of
genus
$G$ and area $A$, the free energy
\begin{equation}
{\cal W}  (G, \lambda A, N) =\ln {\cal Z}  (G, \lambda A, N)
\label{eq:connected}
\end{equation}
describes a string theory identical to that described by ${\cal Z}
(G,
\lambda A, N)$, except that the covering spaces (the string
world-sheets) are restricted to be connected spaces.  It is this
connected string theory that we will use to calculate the sum over
covering maps in the first example below.  We can expand the
partition
function and the free energy explicitly in terms of coefficients
$\eta_{g,G}^{n,i}$ and $\omega_{g,G}^{n,i}$,
\begin{equation}
{\cal  Z}  (G,\lambda A, N) = \sum_{g = -\infty}^{\infty}  \sum_{n}
\sum_{i}
\eta_{g,G}^{n,i} \;
e^{- {n \lambda A\over  2}}(\lambda A)^{i} N^{2 - 2g}.
%\label{eq:}
\end{equation}
\begin{equation}
{\cal W}  (G,\lambda A, N) = \sum_{g = -\infty}^{\infty}  \sum_{n}
\sum_{i}
\omega_{g,G}^{n,i} \;
e^{- {n \lambda A\over  2}}(\lambda A)^{i} N^{2 - 2g}.
%\label{eq:}
\end{equation}
We can perform a similar expansion for the partition function and
free
energy in a single  sector,
\begin{equation}
Z^+(G,\lambda A, N) = \sum_{g = -\infty}^{\infty}  \sum_{n} \sum_{i}
\zeta_{g,G}^{n,i} \;
e^{- {n \lambda A\over  2}}(\lambda A)^{i} N^{2 - 2g}.
%\label{eq:pchiral}
\end{equation}
\begin{equation}
W^+(G,\lambda A, N) = \sum_{g = -\infty}^{\infty}  \sum_{n} \sum_{i}
\xi_{g,G}^{n,i} \;
e^{- {n \lambda A\over  2}}(\lambda A)^{i} N^{2 - 2g}.
\label{eq:pchiral}
\end{equation}
The coefficients $\xi_{g,G}^{n,i}$ when $ 2(g - 1)= 2n (G- 1)+i$
correspond to the $n$-fold orientation-preserving covering maps from
a
connected genus $g$ Riemann surface to ${\cal M}$ with $i$ branch
points.  When $i = 0$, these terms count the number of connected
unbranched $n$-fold covers of ${\cal M}$, where each cover counts as
$1/| S_\nu |$.  The coefficients $\zeta_{g,G}^{n,i}$ similarly
correspond to the number of orientation-preserving covering maps from
a possibly disconnected Riemann service to ${\cal M}$.
\vspace{.15in}

\noindent $\bullet$ {\bf $G = 0$}

As a first example, we consider the partition function of the
connected theory (\ref{eq:connected}) on the manifold $S^2$.  In many
ways this is the most complicated case since here there are
contributions of maps of arbitrarily large winding number even for a
fixed power of $1/N$.

In a single sector, the free energy is given by
\begin{equation}
W^+ (0,\lambda A, N) = \int_{ \tilde{\Sigma}_+ (S^2)}  d \nu
\; e^{- \frac{n \lambda A}{2} }
\frac{(-1)^{i} N^{2 -2 g}}{ |S_\nu |} ,
\label{eq:sphere}
\end{equation}
where the covering maps in $\tilde{\Sigma}_+ (S^2 )$ are continuous
maps from connected spaces to $S^2$ that have exactly two
$\Omega$-points.  We will now calculate the contributions to
(\ref{eq:sphere}) for $n =1$ and $n =2$.

When $n =1$, clearly there is only one covering space; this cover has
a single sheet and no singularities.  When there is only a single
sheet, there can be no branch points and no tubes, and the effect of
$\Omega$-points is restricted to the trivial permutation, so the
contribution to $W^+( 0,\lambda A, N)$ from maps with $n = 1$ is
given
by the trivial single sheeted cover with all possible contracted
handles.  We recover the full contribution by summing over all
possible locations of handles;
\begin{equation}
W^+_1 (0,\lambda A, N) = N D_1 (\lambda A, N)=
e^{- \frac{\lambda A}{2} }
\sum_{h}\frac{(\lambda A)^{h}}{2^h\; h!}
N^{2 (1 - h)}=N^2 e^{- \frac{\lambda A}{2} + \frac{\lambda A}{2N^2}},
%\label{eq:}
\end{equation}
where by $D_1(A)$ we denote the $n =1$ contribution to the partition
function on a disc of area $A$.  We can easily verify this expression
in QCD$_2$; the exponent is just $- (\lambda /2N)
C_2({\,\lower0.9pt\vbox{\hrule \hbox{\vrule height 0.2 cm \hskip 0.2
cm \vrule height 0.2 cm}\hrule}\,})$, where the Casimir of the
fundamental representation ({\,\lower0.9pt\vbox{\hrule
\hbox{\vrule height 0.2 cm
\hskip
0.2 cm \vrule height 0.2 cm}\hrule}\,}) is $N-1/N$.

When $n =2$, the cover has two sheets.  The only possible non-trivial
permutation on these sheets is the permutation that switches sheets
($P$); this is the permutation that every branch point must give, and
is the only non-trivial permutation that may be produced by the
$\Omega$-points.  Thus $\Omega_2= 1+P/N $, and $\Omega_2^2= (1+
1/N^2)
+2P/N$.  The total number of  permutations $P$ produced by branch
points or $\Omega$-points must be even, since the covering space has
no boundary. In addition the sheets may be connected by orientation
preserving tubes.  If there are no branch points and the permutation
{}from each $\Omega$-point is trivial, then there must be at least one
tube, or the covering space would be disconnected.  Every covering
space that has no contracted handles has a symmetry factor of $1/2$,
since the two sheets are identical.  Thus, the contribution to the
free energy from maps with $n =2$ is given by,

\begin{eqnarray}
W_2^+ (0,\lambda A, N) &  &= e^{-\lambda A }
\left[\frac{1}{2} \sum_{h}\frac{(\lambda A)^{h}}{h!}N^{-2h}
\right] \nonumber \\
& & \cdot
\left[ \sum_{t}\frac{(\lambda A)^{  t}}{ t!} N^{-2t }
\sum_{i }\frac{(\lambda A)^{2 i }}{(2i)! } N^{2 (2- i)}
\left(1+ \frac{1}{N^2} + (  \frac{2}{N }) ( \frac{-\lambda A}{ (2i
+1)N})
\right)
  - N^{4}   \right]
\nonumber \\
&  &=  e^{-\lambda A }\left[N^2 \left( \frac{1}{2}   -
\frac{1}{2}  \lambda A +
\frac{1}{4}(\lambda A) ^2 \right)+ {\cal O}  (1) \right].
\label{eq:spherepartition}
\end{eqnarray}
The subtraction of the $N^4$ term is to remove the case of no
branchpoints, no
tubes and trivial $\Omega$-points, that is disconnected. It is
easily seen
that this agrees with QCD$_2$  by rewriting the result as
\begin{eqnarray}
\lefteqn{W_2^+ (0,\lambda A, N)  =}  \nonumber\\
& &\left(   {N(N+1) \over 2} \right)^2 e^{-
{\lambda A \over 2 N}    ( 2N +2 -{4\over N}) } +
 \left( {N(N-1) \over 2}\right)^2 e^{- {\lambda A \over 2 N}
(2N -2  -{4\over N})} - {N^4 \over 2}
e^{- {\lambda A \over  N} (N- {1\over N})},
\end{eqnarray}
and recognizing the first term as the contribution of the
representation $\;
\begin{picture}(12, 10)(0,0)
\put(0,5){\line(1,0){12}}
\put(0,5){\line(0,-1){6}}
\put(0,-1){\line(1,0){12}}
\put(12,5){\line(0,-1){6}}
\put(6,5){\line(0,-1){6}}
\end{picture}
\;$
 to the partition function on the sphere,  the second as the
contribution
of $
\begin{picture}(6,20)(0,0)
\put(0,10){\line(1,0){6}}
\put(0,10){\line(0,-1){12}}
\put(0,4){\line(1,0){6}}
\put(0,-2){\line(1,0){6}}
\put(6,10){\line(0,-1){12}}
\end{picture}
$, and the third as the subtraction of the disconnected
parts.

We can combine these results to calculate the contributions to the
coupled free energy ${\cal W}  (0,\lambda A, N)$ from covers with $n
\leq 2$.  The coupled free energy is given by
\begin{eqnarray}
\lefteqn{{\cal W}  (0,\lambda A, N) =} \\
& & 2 ( W^+_1  (0,\lambda A, N)+W^+_2
(0,\lambda
A, N)) \nonumber\\
& &+W^+_1  (0,\lambda A, N)^2 \left[\sum_{\tilde{t}}
\frac{(-\lambda A)^{\tilde{t}}}{\tilde{t}!} N^{-2 \tilde{t}} (1  -
\frac{1}{N^2} )^2 - 1 \right]  + {\cal O}  (e^{- \frac{3
\lambda A}{2} }).
\end{eqnarray}
The term quadratic in $W_1^+$ arises from a  2-fold
cover of $S^2$ consisting of a single-sheeted cover of each
orientation connected by orientation-reversing tubes.  The term
that is subtracted is the disconnected cover  of this type without
tubes.

This procedure can be continued to higher winding number maps. A very
interesting calculation would be to calculate the exact leading order
contribution (order $N^2$) to ${\cal W} (0,\lambda A, N)$ for
arbitrary winding number. Using the methods of \cite{gross} this
calculation can be treated by discrete orthogonal polynomials;
however,
so far this has not yielded a closed form solution.
%\eject

\noindent $\bullet$ {\bf $g  =G=1 $}

The next example is that of maps from the torus onto the torus. For
such maps there can be no branch points, tubes, contracted handles,
or
nontrivial twists. Thus, we must simply count the total number of
topologically distinct covering of a torus by another torus. If we
consider connected maps then we can only have maps of a given
orientation, since orientation reversing tubes would increase $g$.
Thus ${ \cal W}(1,\lambda A)=2 W^+(1,\lambda A) + {\cal O}(N^{-2}) $.

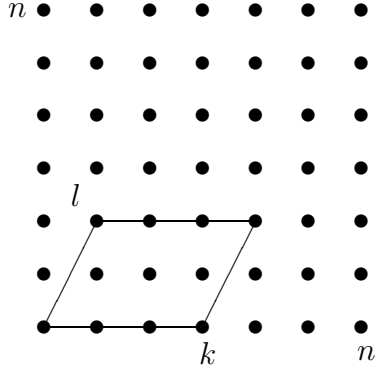
\begin{figure}
\centering
\begin{picture}(200,200)(- 80,- 80)
%\thicklines
\multiput(- 50,- 50)(20,0){7}{\circle*{5}}
\multiput(- 50,- 30)(20,0){7}{\circle*{5}}
\multiput(- 50,- 10)(20,0){7}{\circle*{5}}
\multiput(- 50, 10)(20,0){7}{\circle*{5}}
\multiput(- 50, 30)(20,0){7}{\circle*{5}}
\multiput(- 50, 50)(20,0){7}{\circle*{5}}
\multiput(- 50, 70)(20,0){7}{\circle*{5}}
\put(- 60,70){\makebox(0,0){$n$}}
\put(70,- 60){\makebox(0,0){ $n$}}
\put(- 50,- 50){\line(1,0){60}}
\put( 10,- 60){\makebox(0,0){ $k$}}
\put(- 50,- 50){\line(1,2){20}}
\put(- 30,- 10){\line(1,0){  60}}
\put(10,- 50){\line(1,2){20}}
\put(- 40, 0){\makebox(0,0){ $l$}}
\end{picture}
\caption[x]{\footnotesize A  6-fold map of a torus  with windings 3
and 2  }
\label{f:torusmaps}
\end{figure}

Consider an $n$-sheeted cover of the torus. The two cycles of the
world sheet torus must wind an integral number of times around the
two
cycles of the target space torus; $k$ times around one cycle and $q$
times around the other.  Clearly $n=kq$. To determine how many
topologically inequivalent maps there are with such windings one can
draw a fundamental region of the world sheet torus on the lattice
associated with the universal covering space of the target space
torus.  In Figure~\ref{f:torusmaps} an example is given of a six
sheeted cover with windings 3 and 2. There are many such coverings;
for a given value of $k$ they can be labeled by an integer $l$, where
the world-sheet torus is defined by the vertices $(0,0),(k,0), (l,q),
(k+l, q) $ on the target space lattice.  But the mapping class
transformations of the world sheet can transform $l \to l + mk$.
Therefore the total number of inequivalent tori is
\begin{equation}
 {C_n}=
\sum_{k | n}\sum_{l =0}^{k -1}1  =
\sum_{k | n}k=\sum_{q | n}\frac{n}{q},
\label{eq:count}
\end{equation}
where ${k | n}\Leftrightarrow$ $k$ divides $n$.
The symmetry factor associated with an $n$-fold cover of this type is
always $n$, so the sum of the symmetry factor over all covers is
given
by
\begin{equation}
\xi_{1,1}^{n,0}=
 \sum_{q | n}\frac{1}{q},
\label{eq:symfactor}
\end{equation}
We can now calculate the leading term in $W^+(1,\lambda A) $ as
\begin{equation}
\lim_{N \rightarrow \infty}
W^+(1,\lambda A)= \sum_{n=1}^\infty \sum_{q | n}\frac{1}{q}x^n
=\sum_{m=1}^\infty \sum_{q=1}^\infty
\frac{1}{q}x^{mq}=-\sum_{m=1}^\infty
\ln(1-x^m),
\label{eq:wplus}
\end{equation}
where $x=\exp[-\lambda A/2]$.  Exponentiating the sum of connected
maps then gives us the sum of all maps including disconnected ones.
This should give the leading contribution to the QCD partition
function on a torus as $N\to \infty$.
\begin{equation}
\lim_{N \rightarrow \infty}
Z^+(1,\lambda A) = \prod_m{1 \over (1-x^m)} = \eta(x)=\sum_np(n)x^n,
\label{eq:toruspart}
\end{equation}
where $p(n) $ is the number of partitions of $n$. This is easily
seen to be
identical to the  QCD calculation,
\begin{eqnarray}
 Z^+(1,\lambda A)  &=& \sum_n\sum_{R\in Y_n}e^{ -{\lambda A C_2(R)
\over   2 N }}=
\sum_{n_1\geq n_2\geq \dots   } x^{n_1 +n_2 +\cdots }\left(1+
{\cal O} ({1\over N})\right)\\
& = &\sum_np(n)x^n\left(1+
{\cal O} ({1\over N})\right).
\label{eq:qcdcalcul}
\end{eqnarray}
This calculation  complements the discussion in \cite{previous},
where we
calculated, using arguments along the lines of Section 2, the number
of
inequivalent disconnected maps of a torus onto a torus with  winding
number $n$
(namely $p(n)$.)
\vskip .15truein

\noindent $\bullet$ {\bf $g = 1 + n (G -1)$}

For our next example, we will compute the coefficients
$\zeta_{g,G}^{n,0}$ for arbitrary $G$ when $g = 1 + n (G -1)$.  These
coefficients correspond to a sum over all disconnected
orientation-preserving coverings of
the genus $G$ manifold ${\cal M}$ without branch points,  tubes,
contracted handles, or nontrivial twists.

\begin{figure}
\centering
\begin{picture}(300,200)(- 150,- 100)
\thicklines
\put(0,- 40){\oval(60,60)}
\put(30,- 46){\vector(0,1){8}}
\put(- 110,65){\makebox(0,0){
\begin{picture}(100,100)(- 50,- 50)
\thicklines
\put(0,0){\oval(64,64)}
\put(0,12){\oval(56,32)[t]}
\put(0,-12){\oval(56,32)[b]}
\put(28,- 12){\line(0,1){24}}
\put(-28,- 12){\line(0,1){24}}
\put(28,- 6){\vector(0,1){8}}
\put(32,- 6){\vector(0,1){8}}
\thinlines
\put( 0,- 42){\makebox(0,0){$\sigma_1$}}
\end{picture}
}}
\put(110,65){\makebox(0,0){
\begin{picture}(100,100)(- 50,- 50)
\thicklines
\put(0,0){\oval(64,64)[r]}
\put(0,12){\oval(64,40)[tl]}
\put(0,-12){\oval(64,40)[bl]}
\put(0,12){\oval(56,32)[t]}
\put(0,-12){\oval(56,32)[b]}
\put(28,- 12){\line(0,1){24}}
\put(- 32,12){\line(1,- 6){4}}
\put(- 28,12){\line(-1,-6){2}}
\put(- 32,-12){\line(1,6){2}}
\put(28,- 6){\vector(0,1){8}}
\put(32,- 6){\vector(0,1){8}}
\thinlines
\put( 0,- 42){\makebox(0,0){$\sigma_2$}}
\end{picture}
}}
\put(- 75,30){\vector(1,-1){40}}
\put(75,30){\vector(-1,-1){40}}

\end{picture}
\caption[x]{\footnotesize Elements of $S_2$ describing covers of a
boundary component}
\label{f:2covers}
\end{figure}
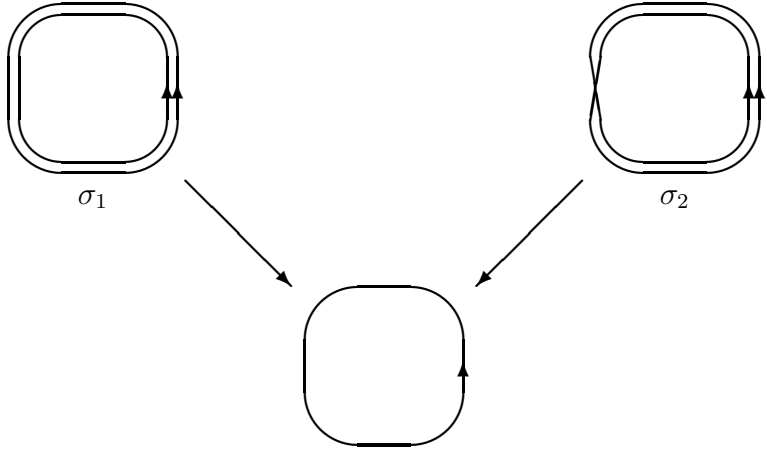

We will compute these coefficients by using the Feynman rules
described in section 4 to construct the manifold ${\cal M}$ out of
spheres with 1, 2 and 3 boundary components (``caps'',
``propagators''
and ``vertices'').  Clearly when $n =1$, $\zeta_{g,G}^{n,0}=1$, since
there is a single cover of any genus Riemann surface with 1 sheet and
no singular points.  We will consider in particular the case $n =2$
while presenting an argument that easily generalizes to arbitrary
$n$.  In order to calculate the sum over double covering spaces of an
arbitrary genus surface from the Feynman diagram point of view, we
will need to know the set of nonsingular double covers of a disk, an
annulus, and a sphere with 3 boundary components.  As in the
algebraic
argument in section 2, in order to count all covers with a factor of
$1/| S_\nu |$, it is simplest to label the sheets of the cover and to
sum a factor of $1/n!$ over all distinct labelings.  The boundaries
of
a 2-fold cover are associated with elements of the symmetric group
$S_2$.  We denote these elements by $\sigma_1,
\sigma_2$, where $\sigma_1$ is the identity element, corresponding to
a disconnected 2-fold cover of the boundary component, and where
$\sigma_2$ is the pairwise exchange corresponding to a connected
2-fold cover (see Figure~\ref{f:2covers}).

In the Feynman diagram picture, we will associate with a disk of area
$A$ a function $D (\sigma, A)$ on $S_n$, giving the sum of $N^n/n!$
over
all labeled covers of the disk with a boundary described by the
permutation $\sigma$.  Similarly, an annulus of area $A$ is described
by a function $P (\sigma,\tau, A)$ where $\sigma$ and $\tau$ are the
permutations describing the boundary components of covering spaces of
the annulus, and a vertex of area $A$ is described by a function $V
(\sigma, \tau, \rho, A)$.  In general these functions contain
covering
maps with arbitrary singularities, and are rather complicated.  Here,
we have restricted to the simpler case where there are no
singularities, so we can write fairly simple expressions for the
functions.  In fact, since the only covers of the sphere without
singularities are given by multiple copies of the trivial 1-fold
cover, it is easy to see that in our case,
\begin{eqnarray}
D (\sigma, A) & = &  \frac{N^n}{n!} \delta (\sigma)
\label{eq:cases1}\\
P (\sigma, \tau, A) & = &  \frac{1}{n!} \delta (\sigma \tau)
\label{eq:cases2}\\
V (\sigma, \tau, \rho, A) & = & \frac{1}{n! N^n}  \delta (\sigma \tau
\rho).
\label{eq:cases3}
\end{eqnarray}
When we glue together these fundamental objects along boundary loops,
we must apply equation (\ref{eq:circleupsilon}).  In terms of the
permutations and labels on the boundaries, this means that when we
glue together an object with boundary $\sigma$ with another object
with boundary $\tau$, we must sum over all possible permutations
$\rho$ describing which labels are identified, and we must insert a
delta function to enforce the condition that under this relabeling
the
permutation is preserved; i.e., that $\rho \sigma \rho^{-1} =
\tau^{-1}$.  (Note that we take the inverse of the permutation
$\tau$,
since the boundary associated with it is of the opposite orientation
{}from the boundary associated with $\sigma$.)
Thus, for instance, we can glue together a disk and a vertex along a
common boundary to reproduce the function $P$ on an annulus,
\begin{equation}
\sum_{\varrho, \rho, \varsigma}V (\sigma,\tau, \rho, A)  D
(\varsigma, B)
\delta (\varrho \varsigma \varrho^{-1}\rho)
= P (\sigma, \tau, A+ B).
\label{eq:annulus}
\end{equation}
Similarly we can glue a disc onto the annulus to reproduce the disc,
\begin{equation}
\sum_{\varrho, \tau, \varsigma}P (\sigma,\tau,  A)  D (\varsigma, B)
\delta (\varrho \varsigma \varrho^{-1}\tau)
= D (\sigma, A+ B),
\label{eq:discglue}
\end{equation}
It is trivial to check that (\ref{eq:annulus}) and
(\ref{eq:discglue}) are
satisfied using (\ref{eq:cases1}), (\ref{eq:cases2}) and
(\ref{eq:cases3}).
Since in the restricted case we are interested in here, none of the
elementary functions are dependent on $A$, we will henceforth drop
the
$A$ dependence of these functions.  In a more general treatment, it
would be necessary to include this dependence in all of the following
equations.

As an intermediate step towards
the result on a surface of arbitrary genus,
we will find it useful to construct the
function $T (\sigma, \tau, A)$ describing a torus with two
boundaries.
\begin{figure}
\centering
\begin{picture}(200,100)(- 100,- 50)
\thicklines
\put(0,12){\oval(32,8)[t]}
\put(0,12){\oval(64,40)[t]}
\put(0,-12){\oval(32,8)[b]}
\put(0,-12){\oval(64,40)[b]}
\put(- 16,- 12){\line(0,1){24}}
\put(16,- 12){\line(0,1){24}}
\put(36,12){\oval(8,8)[bl]}
\put(36,-12){\oval(8,8)[tl]}
\put(-36,12){\oval(8,8)[br]}
\put(-36,-12){\oval(8,8)[tr]}
\put(36,8){\line(1,0){10}}
\put(36,-8){\line(1,0){10}}
\put(-36,8){\line(-1,0){10}}
\put(-36,-8){\line(-1,0){10}}
%\put(46,0){\oval(8,16)[r]}
\put(-46,0){\oval(8,16)}
\thinlines
\put(0,24){\makebox(0,0){
\begin{picture}(10,20)(- 5,- 10)
\thicklines
\put(0,0){\oval(8,16)[ r]}
\thinlines
\put(0,4){\oval(8,8)[tl]}
\put(0,-4){\oval(8,8)[bl]}
\put(- 4,- 2){\line(0,1){4}}
\end{picture}
}}
\put(44,0){\makebox(0,0){
\begin{picture}(10,20)(- 5,- 10)
\thicklines
\put(0,0){\oval(8,16)[ r]}
\thinlines
\put(0,4){\oval(8,8)[tl]}
\put(0,-4){\oval(8,8)[bl]}
\put(- 4,- 2){\line(0,1){4}}
\end{picture}
}}
\put(0,-24){\makebox(0,0){
\begin{picture}(10,20)(- 5,- 10)
\thicklines
\put(0,0){\oval(8,16)[ r]}
\thinlines
\put(0,4){\oval(8,8)[tl]}
\put(0,-4){\oval(8,8)[bl]}
\put(- 4,- 2){\line(0,1){4}}
\end{picture}
}}
\put(60,0){\makebox(0,0){$\tau$}}
\put(-60,0){\makebox(0,0){$\sigma$}}
\put(16,- 24){\makebox(0,0){$A$}}
\end{picture}
\caption[x]{\footnotesize $T (\sigma, \tau, A)$  describes a torus
with two
boundary components}
\label{f:torus}
\end{figure}
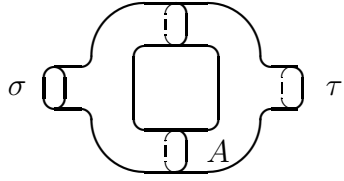
The function $T$ can be constructed by gluing together two vertices
along two boundary loops, and is given by
\begin{eqnarray}
%T (\sigma, \tau) & = & \sum_{\alpha, \beta, \rho, \upsilon, \rho',
%\upsilon'}
%V ( \sigma, \rho, \upsilon) V (\tau^{-1}, \rho', \upsilon')
%\delta (\alpha \rho \alpha^{-1} \rho')
%\delta (\beta \upsilon \beta^{-1} \upsilon') \nonumber\\
% & = & \frac{1}{n!^2} \sum_{\alpha, \beta, \rho}
%\delta (\beta \rho^{-1} \sigma \beta^{-1} \alpha \rho \alpha^{-1}
%\tau^{-1}).
T (\sigma, \tau) & = & \sum_{\alpha, \beta, \rho, \upsilon, \rho',
\upsilon'}
V ( \sigma, \rho, \upsilon) V (\tau, \rho', \upsilon')
\delta (\alpha \rho \alpha^{-1} \rho')
\delta (\beta \upsilon \beta^{-1} \upsilon') \nonumber\\
 & = & \frac{ N^{-2n}}{n!^2} \sum_{\alpha, \beta, \rho}
\delta (\beta \rho^{-1} \sigma^{-1} \beta^{-1} \alpha \rho
\alpha^{-1}
\tau).
\label{eq:handle}
\end{eqnarray}
The sum over
nonsingular covering spaces of a surface of arbitrary genus $G$ can
now be
formed by gluing together $G$ copies of (\ref{eq:handle}), to which
we attach
caps at the two ends.
In general, gluing in a disk to a boundary component $\sigma$ has the
effect of multiplying the function by $N^n \delta (\sigma)$, just as
in (\ref{eq:annulus}) and (\ref{eq:discglue}), since the factor of
$1/n!$ is canceled by the
$n!$  possible ways of gluing the labels together. Thus
 \begin{eqnarray}
\zeta_{g,G}^{n,0} & = & N^{n (2G)} \sum_{\alpha_1, \ldots,
\alpha_{G -1}}
T (\sigma_1,\alpha_1) n! T (\alpha_1, \alpha_2) n!  \ldots n!
T (\alpha_{G-1}, \sigma_1)  \label{eq:intermediate}\\
 & = &  \frac{1}{n!} \sum_{\alpha_1, \rho_1, \ldots, \alpha_{G -1},
\rho_{G -1}} \delta (\prod_{i =1}^{G -1}\alpha_i \rho_i \alpha_i^{-1}
\rho_i^{-1} ) ,
\end{eqnarray}
where in (\ref{eq:intermediate}) we have used the gluing formula
(\ref{eq:circleupsilon}) and the fact that $T (\sigma,\tau)$ is a
class function on $\sigma$ and $\tau$.  This result agrees with
(\ref{eq:result}).

We will now perform this calculation explicitly for the case $n =2$.
In this case, the functions $D$ and $V$ are given by
\begin{eqnarray}
D (\sigma_1)  =    \frac{N^2}{2} ;\ \ \ \
D (\sigma_2)   =    0
\end{eqnarray}
\begin{eqnarray}
V (\sigma_1, \sigma_1, \sigma_1)   =   V (\sigma_1, \sigma_2,
\sigma_2) = \frac{1}{2 N^2} ;  \ \ \ \ \
V (\sigma_1, \sigma_1, \sigma_2)  =   V (\sigma_2, \sigma_2,\sigma_2)
= 0,
\end{eqnarray}
where $V$ is symmetric in its 3 arguments.  The
function $T$ describing a torus with two boundary components is then
given by
\begin{eqnarray}
T (\sigma_1, \sigma_1)   =    T (\sigma_2, \sigma_2)=
\frac{2}{N^4} ; \ \ \ \
T (\sigma_1, \sigma_2)   =    T (\sigma_2, \sigma_1)=  0.
\end{eqnarray}
This function can be viewed as a matrix,  $<\sigma|{\bf
T}|\tau> = N^{2n}T
(\sigma, \tau)\sqrt{| T_\sigma | | T_\tau |}$,  in the space of
conjugacy
classes in $S_2$.  The coefficient $\zeta_{2G -1,G}^{2,0}$ is then
given by
\begin{equation}
\zeta_{2G -1,G}^{2,0} = <\sigma_1|{\bf T}^G|\sigma_1> 2!^{G -1} =
2^{2G -1}.
%\label{eq:}
\end{equation}

Similarly, for arbitrary $n$, we can compute the matrix $<\sigma
|{\bf T}^G|\tau>$ as
a matrix in the space of conjugacy classes in $S_n$.  We then have,
$\zeta_{nG - n +
1,G}^{n,0}=  <\sigma_1 |{\bf T}^G|\sigma_1 >n!^{G-1}$, and we can
compute the coefficient $\zeta_{nG - n +
1,G}^{n,0}$  by decomposing the identity permutation, $|\sigma_1 >$,
into a
linear combination
of eigenvectors of ${\bf T}$.   For $n =3,4$, we get
\begin{eqnarray}
\zeta_{3G -2,G}^{3,0} & = & 2 (6^{2G -2}) + 3^{2G -2}, \\
\zeta_{4G -3,G}^{4,0} & = & 2 (24^{2G -2}) + 2 (8^{2G -2}) + 12^{2G
-2}.
\end{eqnarray}
For general $n$ we note that the normalized eigenvectors of  ${\bf
T}$
are simply  given by the class functions
$ \chi_R(\tau) \sqrt{| T_\tau |}/ \sqrt{n!}$,
with eigenvalues
given by
 ${n! \over d_R^2}$.  This is because
\begin{eqnarray}
\lefteqn{
N^{2n}\sum_{T_\tau}\sqrt{| T_\sigma |}| T_\tau | T(\sigma,
\tau)\chi_R(\tau) } \\
& = &
\sum_{T_\tau}\frac{\sqrt{| T_\sigma |}| T_\tau |}{n!^2}
 \sum_{\alpha, \beta, \rho}
\delta (\beta \rho^{-1} \sigma^{-1} \beta^{-1} \alpha \rho
\alpha^{-1}
\tau) \chi_R(\tau)   \\
& = & \sum_{\alpha, \beta, \rho} \sum_{T_\tau}
\frac{\sqrt{| T_\sigma |}| T_\tau |}{n!^3}\sum_{R'}\chi_{R'}(\beta
\rho^{-1} \sigma^{-1}
\beta^{-1} \alpha \rho \alpha^{-1})\chi_{R'}(\tau) \chi_R(\tau)
\\
& = & \frac{\sqrt{| T_\sigma |}}{n!^2}  \sum_{\alpha, \beta,
\rho}\chi_{R}(\beta
\rho^{-1} \sigma^{-1} \beta^{-1} \alpha \rho \alpha^{-1})    =
\frac{\sqrt{| T_\sigma |}}{n!} \sum_{\alpha, \rho}\chi_{R}(\sigma^{-1}
\alpha  \rho \alpha^{-1} \rho^{-1}) \\
& = & \frac{\sqrt{| T_\sigma |}}{d_R}\sum_{\rho}\chi_{R}(
\alpha  \rho \alpha^{-1}\rho^{-1} ) \chi_{R}(\sigma^{-1})
 = \frac{n!\sqrt{| T_\sigma
|}}{d_R^2}\chi_{R}(\sigma^{-1})
=\frac{n!}{d_R^2}\sqrt{| T_\sigma |}\chi_{R}(\sigma).
\end{eqnarray}
The identity has the decomposition, $\delta(\sigma) =
\delta (\sigma) \sqrt{| T_\sigma |}
=\sum_R \sqrt{| T_\sigma |}
\chi_R( \sigma)
{d_R\over n!}$. Therefore,
\begin{equation}
\zeta_{nG -n+1,G}^{n,0}= \sum_R \left({n!\over d_R^2}\right)^G
(n!)^{G-1}
\left({d_R
\over \sqrt{n!}}\right)^2
= \sum_R ({n!\over d_R})^{2G-2}.
\label{eq:expallp}
\end{equation}

These results are equivalent to those given earlier from the algebraic
point of view; however, this method of calculation gives a better
picture of how these covering maps are described geometrically. We see
that in the case of the torus, $G=1$, we recover again the number of
possibly disconnected maps of the torus onto the torus with winding
number $n$, namely $\zeta_{1,1}^{n,0}=\sum_R 1 = p(n)$.

We can easily check this result in QCD$_2$, since for these maps we
need only  the leading term in the $1/N$ expansion of the dimension
and
the quadratic Casimir, and thus
\begin{equation}
Z^+ \sim \sum_R \left( {d_R N^n \over n!}\right)^{2-2G} e^{-{n
\lambda A\over 2}},
\label{eq:expal}
\end{equation}
in complete accord with (\ref{eq:expallp}). Also, the fact that the
characters are the eigenvalues of ${\bf T}$ is clear from the QCD
calculation,   (\ref{eq:newpart}),
that $T(\sigma,\tau) = \sum_R {1 \over d_R^2 N^{2n}}
\chi_R(\sigma)\chi_R(\tau)$.

The terms we have computed here correspond to possibly disconnected
orientation-preserving covers without singular points.  The
corresponding terms in the coupled theory are given by
\begin{equation}
\eta_{nG -n+1,G}^{n,0} = \sum_{m = 0}^{n}
\zeta_{mG - m+1,G}^{ m,0}\zeta_{(n - m)G -n + m+1,G}^{n- m,0},
%\label{eq:}
\end{equation}
since each nonsingular covering space can be divided into an
orientation-preserving nonsingular covering space and a nonsingular
orientation-reversing covering space, with no orientation-reversing
tubes connecting the two types of spaces, as such tubes would increase
the genus.

\section{Examples of Wilson loops}
\setcounter{equation}{0}
\baselineskip 18.5pt

We will now proceed to calculate the leading terms in the vacuum
expectation values of some simple Wilson loops.  For all the examples
we consider here, we will assume that the Wilson loop is carrying the
integer $k=1$, so that the observable we are measuring is the trace
of
the holonomy of the gauge field around the loop taken in the
fundamental representation.  We will first compute each VEV in a
single sector of the theory, and then describe how it is different in
the coupled theory.  In general, given a Wilson loop $\gamma$ on the
sphere, if the area $A_i$ of a region $M_i$ is taken to infinity, the
computation of $\langle W_\gamma \rangle$ is equivalent to
computing the expectation value of $\gamma$ on the infinite plane.
In
this limit, the number of sheets in the coverings that contribute to
the VEV is limited, since the number of sheets in the infinite region
must be 0.  For such a Wilson loop, this allows us to compute the
complete expectation value as a sum over a finite number of covering
maps.  The expectation values of Wilson loops on the plane have been
previously considered by Kazakov and Kostov using loop equations
derived by Makeenko and Migdal\cite{kk,mm}.  Using these loop
equation
techniques, Kazakov was able to give a general algorithm for
computing
the VEV of an arbitrary Wilson loop in the plane\cite{kazakov}.  N.
Bralic also
derived an algorithm for exact loop averages on the plane using a
non-Abelian
form of Stokes theorem\cite{bralic}.
In both of these  works  the gauge theory was taken to be $U(N)$.
This
corresponds
to dropping the $U (1)$ term in the quadratic Casimir, and
effectively removes handles and tubes from the theory,  except those
orientation-reversing tubes
produced by $\Omega$-points.  We
compare the results computed using the string theory picture with the
computations of Kazakov and Bralic, and verify that they are the
same.

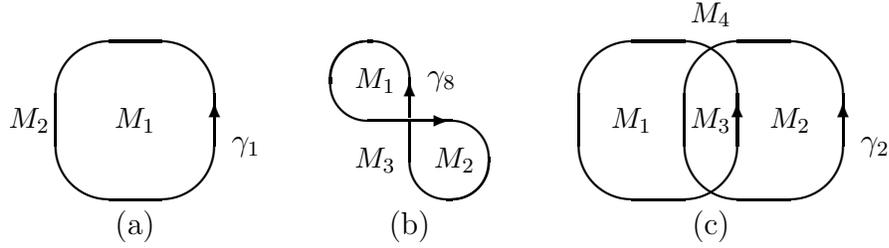
\begin{figure}
\centering
\begin{picture}(100,120)(- 50,- 60)
\put( 0,- 40){\makebox(0,0){(a)}}
\put(0,0){\makebox(0,0){\small $M_1$}}
\put(- 40,0){\makebox(0,0){\small $M_2$}}
\put(40,- 10){\makebox(0,0){ $\gamma_1$}}
\thicklines
\put(0,0){\oval(60,60)}
\put(30,0){\vector(0,1){7}}
\end{picture}
\begin{picture}(110,120)(0,- 60)
\put(35,15){\makebox(0,0){  \small $M_1$}}
\put(35,-15){\makebox(0,0){  \small $M_3$}}
\put(65,-15){\makebox(0,0){  \small $M_2$}}
\put(50,- 40){\makebox(0,0){(b)}}
\put(60,15){\makebox(0,0){ $\gamma_8$}}
\thicklines
\put(35,15){\oval(30,30)[l]}
\put(35,15){\oval(30,30)[t]}
\put(65,-15){\oval(30,30)[b]}
\put(65,-15){\oval(30,30)[r]}
\put(35,0){\vector(1,0){30}}
\put(50,- 15){\line(0,1){14}}
\put(50,1){\vector(0,1){14}}
\end{picture}
\begin{picture}(100,120)(- 50,- 60)
\put( 0,- 40){\makebox(0,0){(c)}}
\put(- 30,0){\makebox(0,0){\small $M_1$}}
\put(30,0){\makebox(0,0){\small $M_2$}}
\put(0,0){\makebox(0,0){\small $M_3$}}
\put(0,40){\makebox(0,0){\small $M_4$}}
\put(60,- 10){\makebox(0,0){ $\gamma_2$}}
\thicklines
\put(- 20,0){\oval(60,60)}
\put( 20,0){\oval(60,60)}
\put(10,0){\vector(0,1){7}}
\put(50,0){\vector(0,1){7}}
\end{picture}
\caption[x]{\footnotesize Wilson loops on the sphere}
\label{f:loops1}
\end{figure}
\vspace{.15in}

\noindent $\bullet$ {$G = 0$, $\gamma =\gamma_1$, $k =1$}

We will first consider the trivial Wilson loop $\gamma =\gamma_1$ on
the
sphere (Figure~\ref{f:loops1}a).  The loop $\gamma$ divides
the sphere into two regions $M_1$, $M_2$ of areas $A_1, A_2$.  The
first step in computing the expectation value of this Wilson loop
in a single sector of the theory
is to assign integers $I_1, I_2$ to the
two regions, indicating how many $\Omega$-points are in each region.
In this case, we have
\begin{equation}
I_1= I_2=1.
%\label{eq:}
\end{equation}
We can now compute a set of relations between the numbers $n_1, n_2$
of sheets over the two regions for any fixed cover.  We will assume
that $M_1$ lies on the inside of $\gamma$.  Since $k =1$, it follows
that for any covering space $n_1= n_2+1$.  The leading term in the
expansion of $\langle W_\gamma \rangle_{+}$ is therefore proportional
to $\exp (- \lambda A_1/2)$,  and corresponds to those covers that
have a single sheet over $M_1$ and nothing over $M_2$.
In fact, there
is only a single such cover, except for possible handles,
since there are no nontrivial permutations on a single sheet.
Suppressing the functional dependence on $\lambda$ and $N$, the first
contribution to $\langle W_\gamma \rangle_{+}$ is given by
\begin{equation}
W_1 (A_1)  = D_1 (A_1)  = e^{- \frac{\lambda
A_1}{2}}\sum_{h}\frac{(\lambda
A_1)^{h}}{2^h\; h!}
N^{1 - 2h}\nonumber   =  N  e^{- \frac{\lambda A_1}{2}
+\frac{\lambda A_1}{2N^2}}.
\label{eq:leading}
\end{equation}

The next set of covering spaces that contribute are
those with
$n_1=2, n_2=1$.  By an argument practically identical to that leading
to (\ref{eq:spherepartition}), the contribution from these covering
spaces is given by
\begin{eqnarray}
W_2 (A_1, A_2)& =&
D_1  (A_2+2 A_1)
%e^{- \frac{\lambda (A_2+2 A_1)}{2}}
%\sum_{h\geq 0}\frac{(\lambda A_2+2A_1)^{h}}{h!}N^{-2h}  \nonumber\\
%& &
\cdot
\sum_{i,t}\frac{(\lambda A_1)^{2 i + t}}{(2i)!\;t!} N^{2- 2i - 2t}
\left(1 - \frac{\lambda A_1}{ (2i +1) N^2})\right). \\
& = & N^2 D_1  (A_2+2 A_1)  e^{  \frac{\lambda A_1}{N^2}}\left[
\cosh{
\frac{\lambda A_1}{N }} -{1\over N} \sinh{ \frac{\lambda
A_1}{N}}\right]
\\  & = & e^{- \frac{\lambda (A_2+2 A_1)}{2}}
\left[N^3+ N \left( \lambda A_1 + \frac{\lambda A_2}{2}  +
\frac{1}{2}(\lambda A_1) ^2 \right)+ {\cal O}  (N^{-1}) \right].
\end{eqnarray}
This equation  is slightly simpler then (\ref{eq:spherepartition})
because here we are working in the space of
possibly disconnected covers, and there is only one
$\Omega$-point in the region $M_1$.

We will now consider the expectation value in the full coupled
theory.
The relation between the numbers $n_i$ of orientation-preserving
sheets and the numbers $\tilde{n}_i$ of orientation-reversing sheets
is
given by
\begin{equation}
n_1- \tilde{n}_1= n_2- \tilde{n}_2 +1,
%\label{eq:}
\end{equation}
where $n_1\geq n_2$ and $\tilde{n}_2\geq \tilde{n}_1$.
The  leading terms in an expansion in the number of sheets are given
by
\begin{eqnarray}
\langle W_\gamma \rangle & = & W_1 (A_1)+ W_1(A_2)
+W_2 (A_1, A_2)+ W_2(A_2, A_1) \nonumber \\
& & +
\sum_{\tilde{t}}
\frac{(-\lambda)^{\tilde{t}}}{  \tilde{t}!}
 N^{-2 \tilde{t}} (1 - \frac{1}{N^2} )
\left[ D_1 (A_2+2 A_1)
 A_1^{\tilde{t}}  \right.\nonumber\\
& &\hspace{1.5in}  \left.+ D_1 (2A_2+ A_1)
A_2^{\tilde{t}}\right]
 + {\cal O}  (e^{- \lambda (A_1+ A_2)})
\end{eqnarray}
where the last term corresponds to the covers with $n_1=
\tilde{n}_1=1$ and $n_2= \tilde{n}_2=1$, connected by an arbitrary
number of orientation-reversing tubes.  Note that the number of
$\Omega$-points in each region is independent of which sector of the
theory we are calculating in this example.

Finally, we consider the VEV of the trivial Wilson loop when the area
$A_2$ goes to infinity.  Since the only covers that contribute in
this limit are those with $n_2 = \tilde{n}_2 = 0$, it follows that
the
only relevant covers are exactly those that gave rise to the leading
term (\ref{eq:leading}), and
\begin{equation}
\lim_{A_2\rightarrow \infty}\langle W_\gamma
\rangle = W_1 (A_1) = N e^{- \frac{\lambda A_1}{2}}
e^{\frac{\lambda A_1}{2N^2}}.
%\label{eq:}
\end{equation}
Here we recognize the exact result of QCD$_2$  for the expectation
value of a
non-intersecting Wilson loop on the infinite plane, $\langle
W_R(A)\rangle
=N^n \exp [-\lambda A C_2(R)/2N]$, for a loop in the fundamental
representation
where $C_2({\,\lower0.9pt\vbox{\hrule \hbox{\vrule height 0.2 cm
\hskip 0.2 cm
\vrule height 0.2 cm}\hrule}\,})=N-1/N$.
\vskip .15truein

\noindent $\bullet$ {$G = 0$, $\gamma =\gamma_8$, $k =1$}

We will next consider the figure-8 Wilson loop $\gamma = \gamma_8$ on
the sphere (Figure~\ref{f:loops1}b).  This loop divides the sphere
into 3 regions $M_1$, $M_2$, $M_3$ of areas $A_1, A_2, A_3$.
This Wilson loop is equivalent
to a Wilson loop winding twice around a point; as  $A_3 \rightarrow
0$, the VEV of this Wilson loop is equal to the VEV of $\gamma_1$
carrying
the integer $k =2$ and the permutation $\sigma_2$.  We will assume
that $M_1$ lies on the inside of $\gamma$, and $M_2$ lies on the
outside.
In the orientation-preserving sector, the numbers of $\Omega$-points
in the 3 regions are given by
\begin{equation}
I_1 = I_2 =1, \; \;I_3= 0.
%\label{eq:}
\end{equation}
The relations between the
numbers of sheets in the 3 regions are
\begin{equation}
n_1= n_3 +1 = n_2 +2.
%\label{eq:}
\end{equation}
In this sector, the leading term in
the expansion of $\langle W_\gamma \rangle_{+}$ is proportional to
$\exp (- \lambda (2 A_1 + A_3)/2)$.
This term is the only one that survives in the limit $A_2\rightarrow
\infty$, and is given by
\begin{eqnarray}
\lim_{A_2\rightarrow \infty}\langle W_\gamma \rangle  & = &
%e^{- \frac{\lambda (2A_1 + A_3)}{2}}
%\sum_{h\geq 0}\frac{(\lambda (2A_1 + A_3))^{h}}{h!}N^{-2h}
%\nonumber\\
D_1  (2A_1 + A_3)
%& &
\cdot
\sum_{i,t}\frac{(\lambda A_1)^{2 i + t}}{(2i)!\;t!} N^{- 2i - 2t}
\left(1-\frac{\lambda A_1}{(2i +1)}   \right)
 \\
&=&
 {1\over 2}D_1  (2A_1 + A_3)\cdot e^{\lambda A_1 \over N^2} [(N+1)
e^{-{\lambda
A_1 \over N }}- (N-1) e^{ {\lambda A_1 \over N }}]
\label{eq:result8}\\
 & = &e^{- \frac{\lambda (2A_1 + A_3)}{2}}\left[  N (1-\lambda A_1 )
+
{\cal O}  (N^0)\right] .
\end{eqnarray}
The leading terms in this expansion correspond to maps from the disk
to $S^2$ with a single branch point in the region $M_1$ arising
either
as a branch point or from the $\Omega$-point in that region.
The expression (\ref{eq:result8}) is identical to the result for the
planar Wilson loop
VEV computed by Kazakov in \cite{kazakov} and by Bralic in
\cite{bralic}, if we
neglect the terms
arising from the
different gauge group.

\begin{figure}
\centering
\begin{picture}(200,100)(- 100,- 50)
\thicklines
\put(- 50,0){\oval(60,60)}
\put(-20,0){\vector(0,1){6}}
\put(-80,0){\vector(0,-1){6}}
\put(35,15){\oval(30,30)[l]}
\put(35,15){\oval(30,30)[t]}
\put(65,-15){\oval(30,30)[b]}
\put(65,-15){\oval(30,30)[r]}
\put(35,0){\vector(1,0){30}}
\put(50,- 15){\line(0,1){14}}
\put(50,1){\vector(0,1){14}}
\put(90,- 15){\makebox(0,0){$\gamma_8$}}
\thinlines
\put(- 15,0){\vector(1,0){30}}
\put(- 75,15){\vector(1,1){10}}
\put(- 75,-5){\vector(1,1){30}}
\put(- 72,-22){\vector(1,1){44}}
\put(- 55,- 25){\vector(1,1){30}}
\put(- 35,-25){\vector(1,1){10}}
\put(25,5){\vector(1,1){20}}
\put(25,15){\vector(1,1){10}}
\put(35,5){\vector(1,1){10}}
\put(75,- 5){\vector(-1,-1){20}}
\put(75,- 15){\vector(-1,-1){10}}
\put(65,- 5){\vector(-1,-1){10}}

\end{picture}

\caption[x]{\footnotesize Map giving leading term in figure 8 VEV}
\label{f:figure8}
\end{figure}
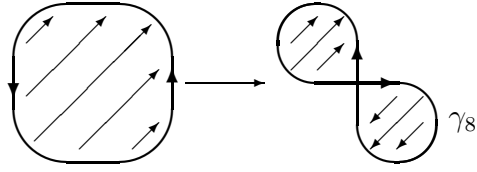

In the coupled theory, the relations between the numbers of sheets
are
given by
\begin{equation}
n_1- \tilde{n}_1= n_3- \tilde{n}_3 +1 = n_2- \tilde{n}_2 +2,
%\label{eq:}
\end{equation}
\begin{equation}
n_1\geq n_3 \geq n_2,\;\; \tilde{n}_2\geq \tilde{n}_3 \geq
\tilde{n}_1.
\label{eq:}
\end{equation}
The first set of covers that contribute are those with $n_3=
\tilde{n}_3=0$ and  $n_1=1,
\tilde{n}_2=1$.  This is also the only set of terms that survive
in the limit $A_3 \rightarrow \infty$.  For these covers, as mentioned
in in section 5, we must take care with the placement of
$\Omega$-points.  The two edges of the graph of $\gamma$ contribute
$\Omega^{-1}$-points to regions $M_1$ and $M_2$.  The set of covers at
the vertex is exactly that described explicitly in the first example
at the end of section 5 (\ref{eq:vertex1}).  The analysis there
indicated that the effect of the vertex was simply to allow the two
sheets of opposite orientation to connect with a twist as part of the
same sheet.  The contributions from these covering spaces are thus
given by
\begin{eqnarray}
\lim_{A_3\rightarrow \infty}\langle W_\gamma \rangle  & = &
D_1 (A_1+ A_2)
\label{eq:8}.
\end{eqnarray}
The leading term in this expansion  corresponds to a map from the
disk
into $M_1\cup M_2$ which is twisted at the intersection point, and
has no other singularities (see Figure~\ref{f:figure8}).
Again, this is exactly consistent with the results given by Kazakov
and Bralic
in
\cite{kazakov}, \cite{bralic}.

Note that in the limit $A_3 \rightarrow \infty$ the VEV
of this Wilson loop
goes to 0 in a single sector, since there are no
orientation-preserving covering maps with $\gamma$ as a boundary and
with no sheets over $M_3$.
\vspace{.15in}

\noindent $\bullet$ {$G = 0$, $\gamma =\gamma_2$, $k =1$}

We now consider the VEV of the pair of interlocking Wilson loops
$\gamma = \gamma_2$ on the sphere (Figure~\ref{f:loops1}c).
In a single sector, the regions
$M_1, M_2, M_3, M_4$ with areas $A_1, A_2, A_3, A_4$ have the
quantities of $\Omega$-points
\begin{equation}
I_1 = I_2 = 0,\; \; I_3 = I_4 = 1.
%\label{eq:}
\end{equation}
Assuming that region $M_3$ is inside of both curves,
the numbers of sheets in each region satisfy
\begin{equation}
n_3= n_1+1 = n_2 + 1 = n_4+2.
%\label{eq:}
\end{equation}
The term corresponding to the covers with $n_3 =2$ is the only term
in the coupled theory
that survives in the limit $A_4 \rightarrow \infty$, and is given by
\begin{eqnarray}
\lim_{A_4\rightarrow \infty}\langle W_\gamma \rangle  & = &
%e^{- \frac{\lambda (A_1 + A_2 + 2A_3)}{2}}
%\sum_{h\geq 0}\frac{(\lambda (A_1 +
%A_2+2A_3))^{h}}{h!}N^{-2h}\nonumber\\
% & &\cdot
D_1 (A_1 + A_2 + 2A_3)
\left[
\sum_{i,t}\frac{(\lambda A_3)^{2 i + t}}{(2i)!\;t!} N^{1- 2i - 2t}
\left(1- \frac{\lambda A_3}{ (2i +1)N^2})\right)
  \right]   \\
& = & N D_1  (A_1 + A_2+2 A_3)
  e^{   \frac{\lambda A_3}{N^2}}\left[
\cosh{
\frac{\lambda A_3}{N }} -{1\over N} \sinh{ \frac{\lambda
A_3}{N}}\right]
\\
 & = & e^{-\lambda (A_1 + A_2+2A_3) }\!\! \left[N^2 \!\!
+\!\! \frac{1}{2}
 \left(\lambda(A_1 + A_2+2A_3) +
(\lambda A_3)^2 \right)+ \! {\cal O}  ({1 \over  N^{2}}) \right],
\end{eqnarray}
in agreement with the calculation of Bralic \cite{bralic}.

In the coupled theory, we can enumerate the covers that contribute in
the limit $A_2 \rightarrow \infty$.  There are two types of covers
that are of interest, each with $n_3=1, \tilde{n}_4 =1$;
one class of covers has $n_1 = \tilde{n}_1=0$, and the other class
has $n_1 = \tilde{n}_1=1$.  The first class of maps has two vertex
points with twists described by (\ref{eq:vertex1}), and only has
handles as possible singularities.  The contribution from this class
of maps is given by
\begin{equation}
\frac{1}{N}  D_1 (A_3+ A_4).
%\label{eq:}
\end{equation}
The leading term here corresponds to a map from the annulus onto $M_3
\cup M_4$ with twists at both intersection points.
The second class of maps has $\Omega$-points distributed according to
the distribution of sheets by the same rules as in a single sector.
The numbers of $\Omega$-points for such covering maps are
\begin{equation}
I_1 = I_2 = 1,\; \; I_3 = I_4 = 0.
%\label{eq:}
\end{equation}
The contribution from maps of this type is
\begin{equation}
D_1 (A_1 + A_3) D_1 (A_1 + A_4) (1 - \frac{1}{N^2} )
\sum_{\tilde{t}}\frac{( -\lambda A_1)^{\tilde{t}}}{\tilde{t}!}
N^{-2 \tilde{t}}.
%\label{eq:}
\end{equation}
The leading term in this contribution arises from a disconnected pair
of sheets; an orientation-preserving sheet over $M_1$ and $M_3$, and
an orientation-reversing sheet over $M_2$ and $M_4$.
The total VEV of the interlocked loops in the limit $A_2 \rightarrow
\infty$ is thus given by
\begin{eqnarray}
\lim_{A_2 \rightarrow \infty}\langle W_\gamma \rangle  & = &
\frac{1}{N}  D_1 (A_3+ A_4) +
N D_1 (2A_1 + A_3 + A_4) (1 - \frac{1}{N^2} )
e^{- \frac{\lambda A_1}{N^2}} \\
& = & N^2 + (\frac{A_3 + A_4}{2} )+ {\cal O}(N^{-2}).
\end{eqnarray}
Neglecting the extra $ U(1)$ terms, this is in agreement with
Kazakov's results in \cite{kazakov}.
\vspace{.15in}

\noindent $\bullet$ {$G = 0$, $\gamma =\gamma_x$, $k =1$}

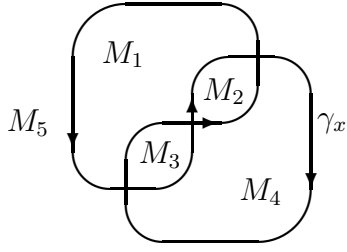
\begin{figure}
\centering
\begin{picture}(200,100)(- 100,- 50)
\thicklines
\put(- 10,10){\oval(70,70)[tl]}
\put(10,30){\oval(30,30)[tr]}
\put(- 30, - 10){\oval(30,30)[bl]}
\put(10, -10){\oval(70,70)[br]}
\put(30,10){\oval(30,30)[tr]}
\put(- 10, - 30){\oval(30,30)[bl]}
\put(10,15){\oval(30,30)[br]}
\put(15,10){\oval(30,30)[tl]}
\put(-10,-15){\oval(30,30)[tl]}
\put(-15,-10){\oval(30,30)[br]}
\put(- 10,0){\vector(1,0){ 20}}
\put(0,- 10){\vector(0,1){ 20}}
\put(25,15){\line(0,1){ 15}}
\put(15,25){\line(1,0){ 15}}
\put(-25,-15){\line(0,-1){ 15}}
\put(-15,-25){\line(-1,0){ 15}}
\put(- 45,25){\vector(0,-1){ 35}}
\put(45,10){\vector(0,-1){35}}
\put(- 10,-45){\line(1,0){35}}
\put( 10,45){\line(-1,0){ 35}}
\put(53,0){\makebox(0,0){$\gamma_x$}}
\put(12,12){\makebox(0,0){$M_2$}}
\put(-12,-12){\makebox(0,0){$M_3$}}
\put(- 26,26){\makebox(0,0){$M_1$}}
\put( 26,-26){\makebox(0,0){$M_4$}}
\put(- 62,0){\makebox(0,0){$M_5$}}
\end{picture}
\caption[x]{\footnotesize Another Wilson loop, $\gamma_x$}
\label{f:loopx}
\end{figure}
We will consider one more Wilson loop on the sphere, the loop $\gamma
=\gamma_x$ shown in Figure~\ref{f:loopx}.  We will only consider the
VEV of this Wilson loop in the planar limit as $A_5 \rightarrow
\infty$.  In this limit, there are no orientation-preserving covering
maps; however, there are 4 classes of coverings which contribute to
the
VEV in the full theory.  In all the relevant cases, the coverings of
$M_1$ and $M_4$ consist of a single orientation-preserving sheet and a
single orientation-reversing sheet, respectively.  The numbers of
sheets in regions 2 and 3 satisfy $n_2 = \tilde{n}_2 \leq 1$, $n_3 =
\tilde{n}_3 \leq 1$.  The four possible cases are described by taking
$( n_2, n_3)=$ $(0,0), (0,1), (1,0), (1,1)$.

In the first case, $( n_2, n_3)= (0,0)$, all three vertices are of the
type described by (\ref{eq:vertex1}).  Only handle singularities are
allowed, so the contribution from this case to the VEV is given by
\begin{equation}
\frac{1}{N^2}  D_1 (A_1+ A_4),
%\label{eq:}
\end{equation}
where the leading term $1/N$ arises from a map from a torus with a
single boundary component onto regions $M_1,M_4$ with twists at all
vertices.

In the cases $( n_2, n_3)= (0,1)$, $( n_2, n_3)= (1,0)$, one of the
two vertices is of the type described by (\ref{eq:vertex1}).  The
other two vertices have $\Omega$-points associated with an adjacent
region with two sheets ($M_3,M_2$ respectively).  The contributions
{}from these two classes of maps are
\begin{equation}
D_1 (A_1 + 2A_3 + A_4) (1 - \frac{1}{N^2} )
\sum_{\tilde{t}}\frac{( -\lambda A_3)^{\tilde{t}}}{\tilde{t}!}
N^{-2 \tilde{t}},
%\label{eq:}
\end{equation}
\begin{equation}
D_1 (A_1 + 2A_2 + A_4) (1 - \frac{1}{N^2} )
\sum_{\tilde{t}}\frac{( -\lambda A_2)^{\tilde{t}}}{\tilde{t}!}
N^{-2 \tilde{t}},
%\label{eq:}
\end{equation}
respectively.

Finally, we consider the case $( n_2, n_3)= (1,1)$.  In this case, the
$\Omega$-points from the outer vertices can be associated with regions
$M_2,M_3$.  The central vertex, however, is of the type described by
(\ref{eq:vertex2}), and carries a factor of $- N (1 -1/N^2)$.  The
contribution from this class of coverings is thus given by
\begin{equation}
-D_1 (A_1 + 2A_2+2A_3 + A_4) (1 - \frac{1}{N^2} )
\sum_{\tilde{t}}\frac{( -\lambda (A_2 +A_3))^{\tilde{t}}}{\tilde{t}!}
N^{-2 \tilde{t}}.
%\label{eq:}
\end{equation}
The total VEV of $\gamma$ in the limit $A_5 \rightarrow \infty$ is
thus given by
\begin{eqnarray}
\lim_{A_5 \rightarrow \infty}\langle W_\gamma \rangle  & = &
\frac{1}{N^2}  D_1 (A_1+ A_4)+
D_1 (A_1 + 2A_3 + A_4) (1 - \frac{1}{N^2} )
e^{- \frac{\lambda A_3}{N^2}}   \nonumber\\
& & +D_1 (A_1 + 2A_2 + A_4) (1 - \frac{1}{N^2} )
e^{- \frac{\lambda A_2}{N^2}}   \\
& &-D_1 (A_1 + 2A_2+2A_3 + A_4) (1 - \frac{1}{N^2} )
e^{- \frac{\lambda (A_2 + A_3)}{N^2}} \nonumber \\
& = & N + \frac{A_1 + A_4}{2N}
+ {\cal O}  (N^{- 3}).
\end{eqnarray}
This agrees again with Kazakov's results \cite{kazakov}  when $U (1)$
terms are neglected.
\vspace{.15in}

\noindent $\bullet$ {$G = 1$, $\gamma =\gamma_a \cup \gamma_{a'}$,
$k =1$}

\begin{figure}
\centering
\begin{picture}(200,100)(- 100,- 50)
\thicklines
\put(0,0){\oval(70,50)}
\put(0,3){\oval(34,20)[b]}
\put(0,0){\oval(30,14)[t]}
\thinlines
\put(0,16){\makebox(0,0){
\begin{picture}(10,20)(- 5,- 10)
\put(0,5){\oval(8,8)[tl]}
\put(0,-5){\oval(8,8)[bl]}
\put(0,0){\oval(8,18)[r]}
\put(- 4,- 2){\line(0,1){4}}
\put(4,-2){\vector(0,1){4}}
\end{picture}
}}
\put(0,35){\makebox(0,0){$\gamma_a$}}
\put(- 7,- 35){\makebox(0,0){$\gamma_{a'}$}}
\put(24,- 8){\makebox(0,0){ $M_2$}}
\put(-24,- 8){\makebox(0,0){ $M_1$}}
\put(- 7,-16){\makebox(0,0){
\begin{picture}(10,20)(- 5,- 10)
\put(0,5){\oval(8,8)[tl]}
\put(0,-5){\oval(8,8)[bl]}
\put(0,0){\oval(8,18)[r]}
\put(- 4,- 2){\line(0,1){4}}
\put(4,2){\vector(0,-1){4}}
\end{picture}
}}
\end{picture}
\caption[x]{\footnotesize Wilson loops on a torus}
\label{f:torusloop}
\end{figure}
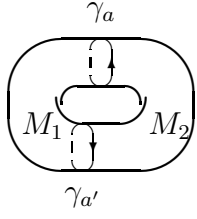

Finally, we will consider an interesting Wilson loop on the torus.  If
we consider a single Wilson loop $\gamma_a$ on the torus that is
homologically nontrivial (see Figure~\ref{f:torusloop}), clearly the
VEV of that Wilson loop must be 0, since the number of sheets of any
cover must differ by 1 across the loop, and the same region is on both
sides of the loop.  The partition function on the torus is physically
equivalent to the finite temperature partition function in flat space.
In the finite temperature case we would regard the length of one cycle
as the spatial extent of the world and the other as $1/\beta $, where
$\beta=k\cdot {\rm Temperature}$. The loop can be thought of as the
Polyakov loop, which is the order parameter for the finite temperature
phase transition. The fact that it always vanishes simply means that
in QCD$_2$ there is no deconfinement transition, no matter how large
the temperature is.

 On the
other hand, if we consider a Wilson loop composed of two disjoint
Wilson loops $\gamma =\gamma_a \cup \gamma_{a'}$ the VEV can
be nontrivial as long as the loops $\gamma_a$ and $\gamma_{a'}$ are
inverses in homology.  If the regions into which the torus is divided
are $M_1, M_2$ then the leading terms in the coupled theory are given
by
\begin{equation}
\langle W_\gamma \rangle = \frac{1}{N}( D (A_1)+ D (A_2)) +
{\cal O} (e^{- \frac{\lambda (A_1 + A_2)}{2}}),
%\label{eq:}
\end{equation}
where the leading terms are given by maps from the annulus into one of
the two regions bounded by $\gamma$.  The structure of this VEV can be
used to analyze the spectrum of heavy quark mesons in the theory.

\section {Dynamical quarks}
\setcounter{equation}{0}
\baselineskip 18.5pt

So far we have described pure QCD$_2$, including an arbitrary number
of Wilson loops, in terms of a geometrical sum of maps--a string
theory. This   allows us to discuss
infinitely heavy quarks in terms of a simple open string theory. More
interesting is to consider the theory with dynamical quarks. This
should be describable in terms of an open string theory with
dynamical boundaries.

In order to relate the full theory to a string theory start with the
partition function for QCD with, for simplicity, a single Dirac
fermion in the fundamental representation of $SU(N)$,
\begin{equation}
Z= \int {\cal D} \bar \Psi {\cal D} \Psi {\cal D} A_\mu e^{ \int d^2
x   [\bar \Psi \gamma^\mu (  \partial_\mu - {\bf A}_\mu)\Psi  -m \bar
\Psi   \Psi - {1\over 4 {\tilde g}^2}{\rm Tr  }F^{\mu \nu}  F_{\mu
\nu} ]}.
\end{equation}
Integrating out the quarks one
is left with
\begin{equation}
Z= \int   {\cal D} A_\mu e^{ -{1\over 4 {\tilde g}^2}\int d^2  x   [
{\rm Tr}  F^{\mu \nu}  F_{\mu \nu} ] +
  {\rm Tr} \ln[\gamma^\mu ( \partial_\mu - {\bf A}_\mu)-m ]}.
\label{eq:partitionquark}
\end{equation}
One can then rewrite the logarithm of the fermion determinant in terms
of a sum over paths using standard techniques. One expresses the
determinant as an integral over proper time ($T$) of the exponential
of the Hamiltonian $H= \gamma^\mu D_\mu-m$, where $D_\mu$ is the
covariant derivative in the background gauge field.  The
usual Hamiltonian path integral representation follows,
\begin{eqnarray}
S_ \Psi(A)&\equiv & {\rm Tr} \ln[\gamma^\mu D_\mu-m ]= {\rm
Tr}\int_0^\infty {dT\over T}
e^{iT(\gamma^\mu D_\mu -m )} \\ &= &  \int_0^\infty {dT\over
T}\oint_{x^\mu(0)=x^\mu(T)}{\cal D} x^\mu(t)   \int {\cal D}p^\mu(t)
{\rm tr}\  P \{  e^{-i \int_o^T\left(  p^\mu \dot x_\mu  -\gamma^\mu
p_\mu +\gamma^\mu {\bf A}_\mu +m \right) dt} \},
\end{eqnarray}
where the  $P$ denotes  path ordering of the Dirac matrices as well
as of the gauge  field (${\bf A}_\mu$), which is a matrix in the
fundamental representation. After    a shift
($p_\mu \to  p_\mu +A_\mu$), we have,
 \begin{equation}
S_ \Psi(A)=  \int_0^\infty {dT\over T}\oint_{x^\mu(0)=x^\mu(T)}{\cal
D} x^\mu(t)        W_x(A) \int {\cal D}p^\mu(t)
{\rm tr}\  P \{  e^{-i \int_o^T\left(  p^\mu \dot x_\mu  -\gamma^\mu
p_\mu  +m \right) dt} \},
\end{equation}
where $W_x(A)= {\rm Tr}P \exp\oint d x^\mu {\bf A_\mu(x)} $ is the
Wilson loop associated with the quark path.

This approach  was studied by Strominger  in 1981 \cite{strominger}.
He   showed that in  two dimensions the  path integral  over the
momentum
$p^\mu$ can be easily evaluated, and that it simply yields a factor
of
$(-1)^{\omega_x } \exp( -imT) $, where $\omega_x $ is the  {\em
winding number}
of the loop, namely the number of $2\pi$ rotations the quark makes
around the closed path. Thus, following Strominger, we see that the
QCD$_2$ partition function can be written as
\begin{equation}
Z= \sum_n{1\over n!}  \prod_{i=1}^n  \left\{   \int_0^\infty
{dT_i\over T_i }\oint_{x^\mu_i(0)=x^\mu_i(T)}{\cal D} x^\mu_i
(-1)^{\omega_{x_i}}
e^{-i m  T_i} \right\} \langle  \prod_{i=1}^n  W_{x_i} (A)
\rangle
\end{equation}

We see that the partition function
can be written as a weighted sum over closed quark paths, times the
expectation value of the Wilson loops in the fundamental
representation for each path. The latter term can be expressed,
following this paper, as a sum of open string maps. (There is, of
course, always the trivial term with no quarks, which corresponds to
closed  string maps). We therefore  can regard the complete partition
function as being given by an open plus a closed string theory, whose
boundaries are dynamical.

If we could find the world sheet string Lagrangian, $ {\cal L}_s$,
that describes the set of maps derived in this paper, we could rewrite
the QCD partition function completely in terms of this string theory;
one of the novel features of this theory would be the fact that the
boundaries of the open string have dynamics of their own.

A challenge for this approach to QCD is to rederive the meson
spectrum calculated by perturbation theory techniques in the large
$N$ limit by 't Hooft \cite{thooft}, \cite{calgro}.  The meson
spectrum is  calculated by finding the poles in correlators of quark
bilinears. These too  have an obvious representation  as sums over
quark paths,
 \begin{equation}
\langle  \bar \Psi \Psi(x)  \bar \Psi\Psi (y)  \rangle =  \int dT
\oint_{ x,y}{\cal D} x^\mu (-1)^{ \omega_{x y}}
e^{-i m  T }  \langle   W_{xy} (A)       \rangle,
\end{equation}
where the sum is over all loops that pass through both $x$ and $y$ in
proper time $T$ and, since $N\to \infty$, we drop all other quark
loops.  Strominger \cite{strominger}, as well as Bars and Hanson
\cite{bars}, argued that this will reproduce  the 't Hooft spectrum.
The argument was that in the infinite momentum frame quarks cannot
backtrack. Thus the quark loops are simple and yield an area law
corresponding to a linear potential. This produces the 't Hooft
spectrum.  However this is not really a string argument. It remains a
challenge to derive the 't Hooft spectrum using the above, manifestly
gauge and Lorentz invariant, formula for the meson correlator, where
one inserts the large $N$ expressions for the Wilson loops.  This is
the procedure that one would follow to calculate the meson spectrum in
the four dimensional string theory of QCD, once the theory has been
constructed.

\section{Conclusions}
\setcounter{equation}{0}
\baselineskip 18.5pt

The primary conclusion of this paper and of our previous work is that
two-dimensional QCD is equivalent to a string theory.  We have shown
that the partition function and the expectation values of Wilson loops
on arbitrary space-time manifolds can be derived in terms of a sum of
maps of a two-dimensional world sheet.  We have given the precise set
of rules that determine the nature of the maps and the weights that
appear in their sum. These rules define the QCD$_2$ string theory.
However, we still do not know a path integral formulation that would
produce this string theory.  In some ways our definition of the QCD
string theory as a sum of maps is more explicit than a specification
of the string Lagrangian.  If one possessed a string Lagrangian, one
would then be faced with the problem of how to perform the path
integral to calculate physical observables. Indeed, one would be happy
to prove that the path integral was equivalent to a simple set of maps
with well defined weights. However, if explicit calculations were our
goal, we might as well stick with QCD in its field theoretic
formulation, where the calculations are even simpler.  Our goals are
different. They are first, to verify that in the simplest of cases QCD
can be described by a string theory, and then to continue the
two-dimensional string theory to four dimensions.

We have succeeded in the first goal--QCD$_2$ is describable by a
string theory. However, the present formulation of the theory does not
allow us to continue to higher dimension. The rules for constructing
the maps that we have presented are inherently two-dimensional and do
not have a straightforward extension to higher dimensions. The most
important feature of the maps is that they exclude all finite folds.
This feature of the string theory, we believe, is responsible for the
absence of all particles in the closed string sector and for the
simplicity of the resulting maps.  In higher dimensions, however,
folds can be removed by an arbitrarily small perturbation of the string
map, and are presumably irrelevant to the theory.  Possibly the
dynamics of the QCD string that eliminates folds in two dimensions
suppresses maps with extrinsic curvature in higher dimensions. We
conjecture that this feature is responsible for the absence of the
tachyon, the dilaton and the graviton. But it is hard to guess the
appropriate formulation of three or four dimensional string theory
with our present understanding. It seems that a necessary intermediate
step is the construction of the first quantized string field theory
that yields our maps.

In constructing the QCD$_2$ Lagrangian we should be guided by the
properties of the maps that we have discovered. The weight of our maps
is the exponential of the area, suggesting that the string action
contains the Nambu-Goto term. Perhaps the most important other feature
of the maps is the suppression of folds of finite size.  This suggests
that the string action contain terms that suppress maps with extrinsic
curvature.  Here we might try to draw some lessons from the recent numerical
simulations of higher dimensional string theories.
Other significant clues are the minus signs associated
with branchpoints and with orientation reversing tubes. This suggests
that the string theory be formulated with some kind of fermionic
variables in addition to the bosonic string coordinates.  Finally, the
$\Omega$ points that occur on target space manifolds of genus not
equal to one and when Wilson loops are present should provide strong
constraints on the string Lagrangian.  The fact that the number of
$\Omega$-points is equal to the Euler characteristic of the surface is
an indication that these objects are related to global structures on
the manifold.
%It seems possible that they are related to mathematical objects such
%as the divisor class of the tangent bundle, whose degree is equal to
%the Euler characteristic, and which might be probed by fermionic
%structures.

In addition to constructing the string action there remains much
interesting work to be done in two dimensions. As we discussed above,
it would be very instructive to reproduce the meson spectrum of
QCD$_2$ (with quarks) by stringy quarks.  It would also be instructive
to reproduce the spectrum of baryons, which should be treated as
solitons in the string theory\cite{kazmig}.

\setcounter{equation}{0}
\baselineskip 18.5pt

\setcounter{section}{0}
\startappendix
\section{Generalized Frobenius relations}
\setcounter{equation}{0}

In this appendix we we generalize the well known Frobenius relations
(\ref{eq:frobenius}) to the case of the composite representations
$\bar S R$.  First let us establish some notations. A conjugacy class
of permutations in the symmetric group $S_n$ on $n$ objects is
determined by the number of cycles of each length in a representative
permutation; all permutations in a conjugacy class have the same set
of cycle lengths.  For a given permutation $\sigma \in S_n$, we denote
the total number of cycles in the permutation by $K_\sigma$.  Given a
permutation $\sigma$ with cycles of length $k_1, \ldots, k_{K_\sigma}$,
we denote the conjugacy class of $\sigma$ by $T_\sigma$.  We will
identify the conjugacy class $T_\sigma$ with the associated set of
cycle lengths $\{k_1, \ldots, k_{K_\sigma}\}$.  Thus, we can apply the
set theoretic operations $\cup,\cap, \subset$ to conjugacy classes.
We will denote by $\sigma_k$ the number of cycles in $\sigma$ of
length $k$.  A standard notation for a permutation is $\sigma= \left[
1^{\sigma_1}\ldots i^{\sigma_i} \ldots n^{\sigma_n}\right]$, where
$\sigma_1+2\sigma_2 + \dots n\sigma_n =n$.  We denote the order of the
conjugacy class of $\sigma$ by $|T_\sigma |$, and the number of
elements of $S_n$ that commute with $\sigma$ by
\begin{equation}
C_\sigma = n!/|T_\sigma | = \prod_{i=1}^n i^{\sigma_i}
 \sigma_i!  .
\label{eq:commuting }
\end{equation}
We will abuse notation slightly by sometimes using conjugacy classes
in place of permutations, where an equation depends only on the
conjugacy class of a permutation.

The standard Frobenius relations are
\begin{eqnarray}
\chi_R (U) & = & \sum_{\sigma\in S_k}\frac{\chi_{R}(\sigma)}{k!}
\Upsilon_{\sigma} (U),  \label{eq:frobenius3} \\
\Upsilon_{\sigma}(U) & = &  \sum_{R\in Y_k} \chi_{R} (\sigma)
\chi_{R} (U),
\label{eq:frobenius4}
\end{eqnarray}
where we recall that
\begin{equation}
\Upsilon_{\sigma} (U) = \prod_{i=1}^{K_\sigma}  ({\rm Tr}\; U^{k_i}).
\label{eq:upsilon}
\end{equation}
We wish to derive a more general set of relations
\begin{eqnarray}
\chi_{\bar{S}R} (U) & = & \sum_{\sigma, \tau}\frac{\chi_{R}(\sigma)
\chi_{S}( \tau)}{n!\tilde{n}!}
\Upsilon_{\bar{\tau}\sigma} (U,U^{\dagger}),   \nonumber\\
\Upsilon_{\bar{\tau}\sigma} (U,U^{\dagger}) & = &  \sum_{R, S}
\chi_{R} (\sigma)
\chi_{S} (\tau)
\chi_{\bar{S}R} (U).
\label{eq:generalfrob}
\end{eqnarray}
We will take these equations to define the class functions
$\Upsilon_{\bar{\tau}\sigma} (U,U^{\dagger})$.
Clearly, for any $\sigma, \tau$ the function
$\Upsilon_{\bar{\tau}\sigma} (U,U^{\dagger})$ can be written as a sum
of terms
proportional to products of the form $\Upsilon_{\sigma'}(U)
\Upsilon_{\tau'}(U^{\dagger})$,  with a leading term of
$\Upsilon_{\sigma}(U)
\Upsilon_{\tau}(U^{\dagger})$.  From (\ref{eq:generalfrob}) and the
orthonormality of the characters,
$\int d U \chi_{\bar{S}R} (U) \chi_{\bar{S}'R'} (U^{\dagger}) =
\delta_{R,R'} \delta_{S,S'} $, it
follows that
\begin{equation}
\int d U \Upsilon_{\bar{\tau}\sigma} (U,U^{\dagger})
\Upsilon_{\bar{\tau}'\sigma'}
(U,U^{\dagger}) = \delta_{T_\sigma, T_{\tau'}} \delta_{T_{\sigma'},
T_{\tau}}
C_\sigma C_\tau.
\label{eq:identity}
\end{equation}
Thus, in order to construct the functions
$\Upsilon_{\bar{\tau}\sigma}
(U,U^{\dagger})$, it will suffice to orthonormalize the  set of
functions
$\Upsilon_{\sigma}(U)
\Upsilon_{\tau}(U^{\dagger})$.

We claim that in fact
\begin{eqnarray}
\Upsilon_{\bar{\tau}\sigma} (U,U^{\dagger}) &=&\sum_{\upsilon \subset
T_\sigma}
\sum_{\upsilon' \subset T_\tau,\upsilon\approx \upsilon'}
(-1)^{K_{\upsilon}} C_\upsilon
\Upsilon_{\sigma \setminus \upsilon} (U)
\Upsilon_{\tau \setminus \upsilon} (U^{\dagger}),
\label{eq:definition}\\
& = &\Upsilon_{\sigma}(U)
\Upsilon_{\tau}(U^{\dagger}) + \cdots
\end{eqnarray}
where $\upsilon$ and $\upsilon'$ are summed over all subsets of the
sets \{$k_1, \ldots, k_{K_\sigma}\}$ and
$\{ k'_1, \ldots, k'_{K_\tau} \}$ that
define
$T_\sigma$, $T_\tau$. Thus $\sigma \setminus \upsilon$ refers to the
class of
permutations whose cycles have lengths belonging to the set
$\{k_1, \ldots, k_{K_\sigma}\} $  minus the set  $\upsilon$.
 We will take (\ref{eq:definition}) as the definition of
$\Upsilon_{\bar{\tau}\sigma} (U,U^{\dagger})$, and prove that
(\ref{eq:identity})
is satisfied.

Every permutation $\sigma$ can be rewritten as a product of
permutations $\sigma^{(l)}$ all of whose cycles are of length $l$.
Formally, we write $\sigma = \prod \sigma^{(l)} $, where
$\sigma^{(l)}=[l^{\sigma_l}]$.  The functions $K_\sigma$ and $C_\sigma$,
defined above, clearly satisfy
\begin{eqnarray}
K_\sigma & = &\sum_{l}K_{\sigma^{(l)}} \\
C_\sigma & = &\prod_{l}C_{\sigma^{(l)}}.
\end{eqnarray}
The functions $\Upsilon_{\sigma}$ can be written as
$\Upsilon_{\sigma}=\prod_{l}\Upsilon_{ \sigma^{(l)}}$.
Similarly, if $\upsilon \subset T_\sigma$, then the set $\sigma
\setminus
\upsilon$ can also be decomposed as $\sigma \setminus
\upsilon=\prod_l(\sigma
\setminus \upsilon)^{(l)}=\prod_l(\sigma^{(l)}\setminus
\upsilon^{(l)}) $.

{}From the Frobenius relations, we know that
\begin{equation}
\int d U \Upsilon_{\sigma} (U) \Upsilon_{\tau}
(U^{\dagger}) = \delta_{T_\sigma, T_{\tau}}
C_\sigma= \prod_{l}\delta_{\sigma^{(l)},
\tau^{(l)}}C_{\sigma^{(l)}}=\prod_{l}\int d U_l
\Upsilon_{\sigma^{(l)}} (U_l)
\Upsilon_{\tau^{(l)}}
(U_l^{\dagger}).
\label{eq:frobprod}
\end{equation}
In other words, in any group integration over products of traces of
$U, U^{\dagger}$, the product over cycles of different lengths can be
taken outside the integral.

We therefore expect that $\Upsilon_{\bar{\tau}\sigma}(U,U^{\dagger})$
can also be written as a product over cycles of equal length, namely
\begin{equation}
\Upsilon_{\bar{\tau}\sigma}(U,U^{\dagger}) = \prod_l \Upsilon_{
\overline{\tau^{(l)}} \sigma^{(l)}}(U,U^{\dagger}) .
\end{equation}
The Ansatz,  (\ref{eq:definition}), is equivalent to the assertion
that
\begin{equation}
\Upsilon_{ \overline{\tau^{(l)}} \sigma^{(l)}}(U,U^{\dagger}) =
\sum_{k=0}^{{\rm min\
}(\sigma_l,\tau_l)} {\sigma_l \choose k} {\tau_l
\choose k}
(-1)^k l^k k!  \left({\rm Tr}\; U^l \right)^{\sigma_l - k}
\left({\rm Tr}\; U^{\dagger l} \right)^{\tau_l - k}.
\end{equation}

It is a consequence of (\ref{eq:frobprod}) that
 \begin{equation}
\int d U \Upsilon_{\bar{\tau}\sigma} (U,U^{\dagger})
\Upsilon_{\bar{\tau}'\sigma'}
(U,U^{\dagger}) = \prod_{l}
\int d U_l \Upsilon_{\overline{\tau^{(l)}}\sigma^{(l)}}
(U_l,U_l^{\dagger})
\Upsilon_{\overline{\tau^{(l)}}'{\sigma^{(l)}}'} (U_l,U_l^{\dagger}).
%\label{eq:}
\end{equation}
 Therefore, our new basis will be orthogonal if   we can prove that
the $\Upsilon_{ \overline{\tau^{(l)}} \sigma^{(l)}}(U,U^{\dagger}) $ are
orthogonal.
Assume then, that we have permutations $ \sigma, \tau, \sigma',
\tau'$
containing $s,t,s', t'$ cycles of length $l$.  Without loss of
generality, we can assume that $s + t \geq s' + t'$, and $s \geq t$.
In order to prove (\ref{eq:identity}) for these permutations, it will
suffice to prove that
\begin{equation}
\int d U \Upsilon_{\bar{\tau}\sigma} (U, U^{\dagger})
\Upsilon_{\sigma'} (U)
\Upsilon_{\tau'}(U^{\dagger}) =  \delta_{s,t'}\delta_{s',t} l^{s
+t}s!
t!.
\label{eq:sufficient}
\end{equation}
It is clear from counting factors of $U$ and $U^{\dagger}$ that the
integral on the left of (\ref{eq:sufficient}) is 0 unless $s + s' = t
+
t'$.  Thus, if we define $r = s - t$, we have $t' = s' + r \leq s$,
$s'
\leq t $.
We can compute (\ref{eq:sufficient}) directly from the definition
(\ref{eq:definition}),
\begin{eqnarray}
\lefteqn{\int d U \Upsilon_{\bar{\tau}\sigma} (U, U^{\dagger})
\Upsilon_{\sigma'} (U)
\Upsilon_{\tau'}(U^{\dagger})} \nonumber\\
& = &
\sum_{i = 0}^{t} \left(\begin{array}{cc}
s \\ i
\end{array}\right)\left(\begin{array}{cc}
t\\ i
\end{array}\right)  (-1)^i  l^i i! \left[
l^{s + s' - i} ( s + s' - i)!  \right]   \label{eq:sum}\\
 & = & \sum_{i=0}^t \frac{s!  t! ( s + s' - i)!  (-1)^i l^{s + s'}
}{(s - i)!  i!  (t - i)!}\\
& = &l^{s + s'}\frac{d^{s'}}{dx^{s'}} |_{x =1} s! x^{r + s'} (x- 1)^t
 = l^{s + s'} \delta_{t,s'} s! t!,
\end{eqnarray}
where in (\ref{eq:sum}) the sum over $i$ runs over the sets  $\upsilon$ that
contain
$i$ cycles of length $l$, chosen in ${s \choose  i}$ different ways.
Thus, we have proven the assertion (\ref{eq:sufficient}).  It follows
that (\ref{eq:identity}) holds for all permutations, and thus that
(\ref{eq:definition}) correctly defines the functions for the
generalized Frobenius relations.

We can also invert this relation and express the products
$\Upsilon_{\sigma'}(U)
\Upsilon_{\tau'}(U^{\dagger})$
in terms of the $\Upsilon_{\bar{\tau}\sigma} (U,U^{\dagger}) $.
Remarkably, the inverse formula has the same form (except for the
absence of the $(-1)^{K_\upsilon}$ factor),
\begin{equation}
\Upsilon_{\sigma}(U)
\Upsilon_{\tau}(U^{\dagger})=\sum_{\upsilon \subset T_\sigma}
\sum_{\upsilon' \subset T_\tau,\upsilon\approx \upsilon'}
C_\upsilon
\Upsilon_{\overline{\tau \setminus \upsilon}\ \sigma \setminus
\upsilon} (U,U^{\dagger}),
\label{eq:inverse}
\end{equation}
This can be proved by using the inverse for the terms with cycles of
equal length,
\begin{equation}
 \Upsilon_{\sigma^{(l)}} (U)
\Upsilon_{\tau^{(l)}}(U^{\dagger})=
\sum_{k=0}^{{\rm min\ }(\sigma_l,\tau_l)} {\sigma_l
\choose k} {\tau_l \choose k}
  l^k k!  \Upsilon_{ \overline{\tau^{(l)}\setminus [l^k]}\
\sigma^{(l)}\setminus [l^k]}(U,U^{\dagger})
\end{equation}

We give some examples of this orthonormal basis:
\begin{eqnarray}
\Upsilon_{\overline{[1]}\,[1]} (U,U^{\dagger})& = & {\rm Tr}(U) \, {\rm
Tr}(U^{\dagger}) -1\\
\Upsilon_{\overline{[1^3 ]}\,[1^1,2^1 ]} (U,U^{\dagger})& = &
{\rm Tr}(U^2){\rm Tr}(U){\rm Tr}(U^{\dagger })^3 -3 {\rm
Tr}(U^{2}){\rm
Tr}(U^{\dagger})^2 \\
\Upsilon_{\overline{[ 1^n]}\,[1^n ]} (U,U ^{\dagger})& = &
\!\!\!\! {\rm Tr}(U )^n \, {\rm Tr}(U^{\dagger })^n +
\!\!\!  \sum_{k=1}^n {n \choose k}^2  (-1)^k k!{\rm Tr}(U)^{n-k} \,
{\rm
Tr}(U^{\dagger })^{n-k} \\
\Upsilon_{\overline{[n ]}\,[n ]} (U, U^{\dagger })& = &
{\rm Tr}(U^n) \, {\rm Tr}(U^{\dagger n }) -n.
\end{eqnarray}

A direct result of (\ref{eq:definition}) is a useful
relation for the dimensions of composite representations,
\begin{equation}
\dim \bar{S}R = \sum_{\sigma, \tau}
\frac{\chi_{R}(\sigma)
\chi_{S}( \tau)}{n!\tilde{n}!}
\sum_{\upsilon \subset T_\sigma}
\sum_{\upsilon' \subset T_\tau,\upsilon\approx \upsilon'}
(-1)^{K_{\upsilon}} C_\upsilon
N^{K_{\sigma \setminus \upsilon}} N^{K_{\tau \setminus \upsilon}}.
\label{eq:compdim}
\end{equation}
Alternatively, if we use the product form,
\begin{equation}
\dim \bar{S}R = \sum_{\sigma, \tau}
\frac{\chi_{R}(\sigma)
\chi_{S}( \tau)}{n!\tilde{n}!}
 N^{K_{\sigma}+{K_{\tau}}} \prod_l F_{\sigma^{(l)}, \tau^{(l)}} \big(
{l\over N^2} \big) ,
\label{eq:compdim2}
\end{equation}
where
\begin{equation}
F_{\sigma, \tau}\big( x \big) \equiv \sum_v (-1)^v  v! {\sigma
\choose v}{\tau \choose v}x^v = (1-x{d \over dy})^\tau  y^\sigma
|_{y=1}.
\end{equation}
These formulas are appropriate for a $1/N$ expansion of the
dimensions.

\vskip .5truein
{\Large{\bf Acknowledgements}}

We would like to thank Orlando Alvarez  for helpful discussions.

\end{document}